\documentclass[a4paper,11pt]{article}
\pdfoutput=1
\usepackage{jcappub}

\usepackage[utf8]{inputenc}
\usepackage[T1]{fontenc}
\usepackage{graphicx,amssymb,amsmath,wasysym}
\usepackage{xcolor}

\def\nm{{\cal X}}
\newcommand{\ba}{\begin{eqnarray}}
\newcommand{\ea}{\end{eqnarray}}
\newcommand{\be}{\begin{equation}}
\newcommand{\ee}{\end{equation}}
\newcommand{\nn}{\nonumber}

\begin{document}

\begin{flushright}
	PI/UAN-2018-621FT
\end{flushright}

\title{Non-minimally Coupled Pseudoscalar Inflaton}

\author[1]{Juan P. Beltrán Almeida}
\emailAdd{juanpbeltran@uan.edu.co}
\author[2]{and Nicolás Bernal}
\emailAdd{nicolas.bernal@uan.edu.co}
\affiliation[1]{Departamento de Física, Universidad Antonio Nariño\\
Carrera 3 Este \# 47A-15, Bogotá DC, Colombia}
\affiliation[2]{Centro de Investigaciones, Universidad Antonio Nariño\\
Carrera 3 Este \# 47A-15, Bogotá DC, Colombia}

\abstract{
We consider a scenario in which the inflaton $\phi$ is a pseudoscalar field non-minimally coupled to gravity through a term of the form $\nm R \phi^2$.  The pseudoscalar is also coupled to a $U(1)$ gauge field (or an  ensemble of ${\cal N}$ gauge fields) through an axial coupling of the form $\phi F \tilde{F}$. After M.~M.~Anber and L.~Sorbo, Phys.\ Rev.\ D {\bf 81}, 043534 (2010), Ref.~\cite{Anber:2009ua}, it is well known that this axial coupling leads to a production of gauge particles which acts as a friction term in the dynamics of the inflaton, producing a slow-roll regime even in presence of a steep potential.  
A remarkable result in this scenario, is that the spectrum of the chiral gravitational waves sourced by the scalar-gauge field interplay can be enhanced due to the non-minimal coupling with gravity, leading to measurable signatures, while maintaining agreement with current observational constraints on $n_s$ and $r$.
The inclusion of non-minimal coupling could be helpful to alleviate tensions with non-Gaussianity bounds in models including axial couplings. 	
}

\maketitle
\section{Introduction}

\label{Intro}

In this work we revisit the scenario in which a pseudoscalar field is responsible for driving inflation. We consider a pseudoscalar inflaton coupled to a set of $\cal{N}$ $U(1)$ gauge fields through an axial coupling term of the form $(\alpha/f)\,\phi\,F\tilde{F}$, where $\tilde{F}$ is the dual of the field strength $F$ of the gauge field $A_{\mu}$. This scenario was studied in Refs.~\cite{Anber:2009ua, Anber:2012du} with a steep natural cosine potential and then in Refs.~\cite{Barnaby:2010vf, Sorbo:2011rz, Barnaby:2011vw} for arbitrary potentials supporting slow-roll evolution.  This model presents very appealing possibilities such as the sourcing of chiral gravitational waves due to the axial coupling \cite{Sorbo:2011rz, Anber:2012du,Barnaby:2010vf, Barnaby:2011vw, Jimenez:2017cdr}, the raising of signatures of parity violating and anisotropic correlations~\cite{Shiraishi:2013kxa,Bartolo:2014hwa}, possible connections with magnetogenesis~\cite{Durrer:2010mq,Caprini:2014mja}, generation of the baryon asymmetry of the Universe~\cite{Jimenez:2017cdr} and production of primordial black holes~\cite{Garcia-Bellido:2016dkw} among others.
Despite of its several appealing features, this mechanism is severely constrained given that the generation of any measurable signature is unavoidably accompanied with the production of a large amount of non-Gaussianities in the correlations~\cite{Barnaby:2010vf, Barnaby:2011vw, Ferreira:2014zia}.
It was also pointed out that the model might suffer from severe perturbativity constraints~\cite{Ferreira:2015omg},
but recent analyses~\cite{Peloso:2016gqs} claim that in the perturbative regime the model still allows for observable signals at CMB scales.\footnote{However, in the case considered here, the backreaction due to the gauge field production is the responsible for the slow-roll evolution; the backreaction constraints are therefore broken by construction. We assume perturbativity of the scalar and tensor perturbations but we consider that the classical background is driven by the non-perturbative gauge field production process.}

A number of studies have focused on different versions of models including axial couplings and have addressed several theoretical and phenomenological aspects of them. An incomplete list of references includes~\cite{Anber:2006xt,Barnaby:2011qe,Barnaby:2012tk,Dimopoulos:2012av,Meerburg:2012id,Barnaby:2012xt, Linde:2012bt, Cook:2013xea,Caprini:2014mja,Fleury:2014qfa,Ferreira:2014zia,Ferreira:2015omg,Bartolo:2015dga, Namba:2015gja, Peloso:2016gqs, Domcke:2016bkh,Shiraishi:2016yun,Garcia-Bellido:2016dkw,Almeida:2017lrq, Caprini:2017vnn}. Very recently, a closely related approach, studying the production of fermions through a derivative coupling with a pseudoscalar inflaton was addressed~\cite{Adshead:2018oaa}.

Moreover, non-minimal couplings to gravity in the context of inflation have a long history. The main interest in some early references on the subject~\cite{Futamase:1987ua,Fakir:1990eg,Makino:1991sg,Muta:1991mw,Mukaigawa:1997nh,Komatsu:1997hv,Komatsu:1999mt} was to avoid (or rephrase) fine tuning problems in chaotic inflationary models. Later, the interest was boosted by the study of Higgs inflation~\cite{Bezrukov:2007ep}, Higgs-dilaton models~\cite{GarciaBellido:2011de} and $\alpha$-attractors \cite{Kallosh:2013hoa,Kallosh:2013maa,Kallosh:2013tua,Kallosh:2013yoa,Racioppi:2018zoy}.  The case of multifield inflationary models with non-minimal couplings to gravity~\cite{Kaiser:2010yu,Greenwood:2012aj,Kaiser:2013sna} was also considered, and more recently there have been several works on non-minimally coupled inflaton in non-metric formulations of gravity~\cite{Tenkanen:2017jih,Markkanen:2017tun,Jarv:2017azx}.

The aim of this work is to understand how the introduction of a non-minimal coupling to gravity affects the dynamics and the inflationary predictions, in models where an axion plays the role of the inflaton. To this end, we focus specifically in the axion inflation model with a naturally steep potential model presented in Refs.~\cite{Anber:2009ua,Anber:2012du} in the presence of non-minimal coupling with gravity. In this context, the possibility of generating dark matter from primordial black holes was recently discussed~\cite{Domcke:2017fix}, in the case of the so-called $T$-models~\cite{Kallosh:2013hoa,Kallosh:2013maa,Kallosh:2013tua,Kallosh:2013yoa} which lead to the class of $\alpha$-attractors.
Complementary to that discussion, here we consider the perturbations in a steep cosine-like potential. We perform a detailed calculation of the spectrum of scalar and tensor perturbations, and we extract the spectral index of the scalar perturbations $n_s$, together with the tensor-to-scalar ratio $r$ in order to contrast  them with current observational bounds. An interesting result in this case is that, instead of a global suppression effect, for certain regions of the parameters space the non-minimal coupling term suppresses vacuum gravitational waves due to the scalar-gravity interaction, but at the same time, it acts as an enhancer for the sourced gravitational waves due to the axial coupling.
The interest in this combined effect is twofold: On one hand, the soured gravitational waves are chiral because of the preferred helicity in this model, a fact that can be seen as a signature of parity violation in the CMB correlations.
Additionally, the predicted tensor-to-scalar ratio could be within the reach of current or planned future experiments such as BICEP3~\cite{Wu:2016hul}, LiteBIRD~\cite{Matsumura:2013aja, Shiraishi:2016yun} and the Simons Observatory~\cite{Ade:2018sbj}.
On the other hand, we want to explore the constraints coming from interactions between the inflaton, massless gauge fields and gravity, leading to particular signatures and predictions.

The paper is organized as follows: In section~\ref{mdyn} we briefly revisit the axion dynamics in the minimally coupled inflationary setup. In section~\ref{perturbations1} we discuss scalar and tensor perturbations and we review the calculation of the corresponding correlators. We calculate $n_s$ and $r$, and we also comment on the constraints coming from non-Gaussianities. In section~\ref{nmcg} we introduce non-minimal coupling with gravity for the pseudoscalar field and study its implications in the correlation functions and other observables.
Then, section~\ref{numnmc} presents our main results, where we scan the parameter space and compare it with Planck measurements of $n_s$ and $r$. Finally, in section~\ref{Conclusions} we end up with our conclusions.


\section{Minimally Coupled Pseudoscalar and Gauge Field Dynamics}\label{mdyn}
 
Here, we briefly summarize the particle production mechanism due to the pseudoscalar-vector coupling. We revisit some details of the amplification of the gauge fields depending on the helicity of the vector mode~\cite{Anber:2006xt,Anber:2009ua,Barnaby:2010vf,Durrer:2010mq,Sorbo:2011rz,Barnaby:2011vw,Barnaby:2011qe,Barnaby:2012xt,Anber:2012du,Cook:2013xea,Barnaby:2012tk,Ferreira:2014zia,Namba:2015gja,Peloso:2016gqs}. 

Starting from the Lagrangian for $ {{\cal N}}$ $U(1)$ vector fields coupled to the axion field $\phi$~\footnote{Throughout this paper the signature $(-,+,+,+)$ is used.}
\ba\label{lsg} 
S &=& \int d^{4}x \sqrt{-g}\left[   \frac{{M_P^{2}}}{2} R -\frac{1}{2}(\partial \phi)^2 - V(\phi) - \frac{{{\cal N}}}{4}F^{\mu \nu}F_{\mu \nu} - \frac{ {{\cal N}}\alpha}{4f} \phi\, F^{\mu \nu}\tilde{F}_{\mu \nu} \right]\,,
\ea
we derive the system of equations of motion for the pseudoscalar and the vector fields
\ba\label{eoms}
&& \Box \phi - V_{\phi} -\frac{ {{\cal N}}\alpha }{4f} F^{\mu \nu}\tilde{F}_{\mu \nu}  = 0\,, \\\label{eoms2}
&& \nabla_{\mu}\left(F^{\mu \nu}  +  \frac{\alpha}{f} \phi\, \tilde{F}^{\mu \nu}\right) = 0\,,
\ea
where $V_{\phi} \equiv \partial V / \partial \phi$.  For the vector field we also add the Bianchi identity $  \nabla_{\mu} \tilde{F}^{\mu \nu} = 0.$
We work with the vector potential $A_{\mu}$ corresponding to the field strength $F_{\mu \nu} \equiv \nabla_{\mu}A_{\nu} - \nabla_{\nu}A_{\mu}$, in the Coulomb gauge $A_0 = \partial_i A_i=0$. We consider the de Sitter (or quasi de Sitter) background metric
\be
d s^2 = -d t^2 + a^2(t) d\vec{x}^2 = a^2(\tau)\left[- d \tau^2 +  d\vec{x}^2\right]\,,
\ee
where $a \approx -1/(H\tau)$ and $\tau$ is the conformal time. The gradient of the pseudoscalar field $\vec\nabla \phi=0$ is neglected, since homogeneous background solutions are considered. All in all, Eqs.~\eqref{eoms} and ~\eqref{eoms2} are reduced to:
\ba\label{phitau}
&& \phi'' + 2\,a\,H\,\phi'  + a^2\, V_{\phi}   = a^2 \,\frac{ {{\cal N}}\alpha}{f} \vec{E}\cdot \vec{B}\,,  \\
\label{Atau}
&& A''_i - \nabla^2 A_i -\frac{\alpha}{f}\phi'\,\epsilon_{ijk} \nabla_{j}A_{k} = 0\,, 
\ea 
where primes denote derivatives with respect to conformal time and the electric and magnetic components are defined as
\be\label{EBdef}
E_{i} = -\frac{1}{a(\tau)^2} F_{0i} = -\frac{1}{a(\tau)^2} \frac{\partial A_i}{\partial \tau} \qquad \mbox{and} \qquad B_{i} = \frac{1}{a(\tau)^2}\tilde{F}_{0i}= \frac{1}{a(\tau)^2} \epsilon_{ijk}\nabla_{j}A_{k}\,.
\ee
With these definitions it is possible to check that 
\be\label{EBid}
F\tilde{F} = -4 \vec{E}\cdot\vec{B}\qquad \mbox{and} \qquad F^2 = {2} (\vec{B}^2 - \vec{E}^2)\,.
\ee
The dynamics of the system is completed with Einstein's equations:
\be\label{Rmunu}
{R}_{\mu\nu} - \frac{1}{2} {g}_{\mu\nu} {R} = \frac{1}{M_P^2} {T}_{\mu\nu}\,,
\ee
where the energy momentum tensor is
\be
{T}_{\mu\nu} = \frac{-2 }{\sqrt{-{g}}} \frac{ \delta {\cal L} }{\delta {g}^{\mu\nu}} =    {\partial}_{\mu} \phi {\partial}_{\nu} \phi - {g}_{\mu\nu}\left( \frac{1}{2}{g}^{\alpha \beta} {\partial}_{\alpha} \phi {\partial}_{\beta} \phi + {V}  \right) + {{\cal N}}{F}_{\mu\alpha} {F}_{\nu}{}^{\alpha}  - {g}_{\mu\nu} \frac{{{\cal N}}}{4 }  {F}^{2}\,.
\ee
The Friedmann equations obtained from the Einstein equations (\ref{Rmunu}) are
\ba
 H^2 & = & \frac{1}{  3M_P^2}\left(\frac{1}{2}\dot{\phi}^2 + V(\phi) + \frac{ {{\cal N}}}{2}(\vec{E}^2 + \vec{B}^2)\right)\,, \\
\dot{{H}}   &=& - \frac{1}{  2M_P^2}   \dot{\phi}^2   - \frac{ {{\cal N}}}{ 3 M_P^2 }   ( \vec{E}^2 + \vec{B}^2 )\,.
\ea
To quantize the system, we write  the vector potential as an operator and decompose it in terms of creation and annihilation operators as
\be
\hat{A}_{i} (\tau, \vec{x}) = \sum_{\lambda = \pm} \int \frac{d^3k}{(2\pi)^{3/2}} \left[ \epsilon^{i}_{\lambda}(\vec{k}) \, A_{\lambda} (\tau, \vec{k})\, \hat{a}_{\lambda} (\vec{k})\, e^{i\vec{k}\cdot \vec{x}} + \mbox{c.c.} \right]\,,
\ee 
where the (transverse) polarization vectors are defined as such:
\be 
\vec{\epsilon}_{\lambda}(\vec{k}) \cdot \vec{k} = 0, \quad   \vec{k} \times \vec{\epsilon}_{\lambda}  = - i \lambda |\vec{k}| \vec{\epsilon}_{\lambda}, \quad \vec{\epsilon}_{\lambda}\cdot \vec{\epsilon}_{\lambda'}  =\delta_{\lambda, -\lambda'},  \quad {\vec{\epsilon}}^{\; *}_{\lambda}(\hat{k})= \vec{\epsilon}_{-\lambda} (\hat{k}) = \vec{\epsilon}_{\lambda} (-\hat{k}) \,.
\ee
Then, the vector equation in momentum space is written as 
\be \label{eqApp}
A''_{\pm}  + \left( k^2  \mp \frac{\alpha \,k}{f} \phi' \right)A_{\pm} = 0\,. 
\ee
During the inflationary regime, the classical background solution of the pseudoscalar field is such that $\frac{d\phi}{dt} = \dot{\phi_0} \approx \mbox{constant}$.
Then
\be
\phi' = \frac{d\phi}{dt} \frac{dt}{d \tau} = -\frac{1}{H\tau} \frac{d\phi}{dt} \approx  -\frac{1}{H\tau} \dot\phi_0
\ee
and Eq.~\eqref{eqApp} becomes
\be \label{apm}  
A''_{\pm}  + \left( k^2  \pm \frac{2 \,k \,\xi}{\tau} \right)A_{\pm} = 0 \qquad \mbox{with} \qquad  \xi \equiv \frac{\alpha\, \dot{\phi_0}}{2\,f\, H}\,.
\ee
Eq.~\eqref{apm} is the starting point for the study of the amplification of the gauge fields. Notice that this equation shows manifestly the parity-violating nature of the system, since one of the modes ($A_{+}$) is amplified,\footnote{Taking $\alpha>0$ and $\xi>0$.} while the other ($A_{-}$) is exponentially suppressed (since $\tau <0$). The solution consistent with the Bunch-Davies vacuum initial conditions  for the regime $|k\,\tau| \ll 2\,\xi$ is~\cite{Anber:2009ua}
\be\label{Asol}
A_{+} \approx \frac{1}{\sqrt{2k}} \left( \frac{k}{2\,\xi\, a\, H}\right)^{1/4} e^{\pi\, \xi - 2 \sqrt{2\xi \,k/(aH)}}\,.
\ee  
Here we notice that the $A_+$ mode is exponentially amplified by the factor $e^{\pi \xi}$. Hence, the size of the backreaction effects due to the presence of the vector field coupling~\cite{Anber:2009ua} are
\be\label{EBxi}
\langle \vec{E}\cdot\vec{B}\rangle \approx -{\cal I}\, \frac{H^4}{\xi^4}\,e^{2\pi \xi}, \,\quad  \frac{1}{2}\langle \vec{E}^2 + \vec{B}^2\rangle \approx \frac{4 {\cal I} }{7}\frac{H^4}{\xi^3}\,e^{2\pi \xi},  \,\quad \mbox{with} \,\quad {\cal I} \approx 2.4\times 10^{-4}\,.
\ee 

\subsection{Perturbations} 
\label{perturbations1}
Now we can write the equations for the perturbations of the system including gravity, the inflaton and the gauge fields. They can be obtained varying Eq.~\eqref{phitau} and taking the decomposition for the inflaton as a  background value plus a perturbation of the field $\phi(\tau, x) = \phi_0 (\tau) + \delta \phi(\tau, x)$:
\ba \label{phi0}
&& \phi''_0 + 2a\,H\,\phi'_0   +  a^2 \,V_{\phi}  =  a^2 \frac{ {{\cal N}}\alpha}{f} \langle \vec{E}\cdot \vec{B} \rangle \,, \\ \label{phipert}
 && \delta\phi'' + 2a\,H\,\delta\phi'  - \nabla^2\delta\phi +  a^2 \,V_{\phi\phi}\, \delta\phi  =  a^2 \frac{ {{\cal N}}\alpha}{f} \delta\left[\vec{E}\cdot \vec{B} \right]\,. 
\ea
The original calculation was performed in Ref.~\cite{Anber:2009ua}; for completeness we revisit all the relevant details in Appendix \ref{AA}. 
For small inhomogeneities the $\nabla^2$ term can be neglected.
We consider also that $V\approx 3M_P^2\, H^2$ and $ V_{\phi} \approx V/f$.
Assuming $\alpha \gg 1$ and $f\lesssim M_P$, which means that the dominant contribution comes from the backreaction term $\delta\left[\vec{E}\cdot \vec{B} \right]$, Eq.~\eqref{phipert} becomes
\be \label{pertphiapp}
\delta\phi'' {+} \frac{1}{\tau}   \frac{\alpha\, \pi\, V_{\phi} (\phi_0)}{f\,H^2}\delta {\phi}' +    \frac{1}{\tau^2} \frac{ V_{\phi\phi}(\phi_0)}{H^2} \delta\phi  \approx  a^2 \frac{ {{\cal N}}\alpha}{f} \delta_{\vec{E}\cdot \vec{B}}\,,
\ee
which, in Fourier space,  can be formally solved as
\be\label{delphi}
\delta \phi(\tau, \vec{k}) =  \frac{ {\cal N} \alpha}{f} \int_{-\infty}^{\tau} d\tau_1 \,a^2(\tau_1)\, G(\tau, \tau_1) \int d^3x\, e^{-i\vec{k}\cdot\vec{x}} \, \delta_{\vec{E}\cdot \vec{B} }(\tau_1, \vec{x}) \,.
\ee
$G(\tau, \tau_1)$ is the solution of the  Green's function associated with Eq.~\eqref{pertphiapp}, this is
\be\label{greentext}
\frac{\partial^2G (\tau ,\tau')}{\partial \tau^2}  {+} \frac{1}{\tau}   \frac{  \alpha\, \pi\, V_{\phi}(\phi_0)}{f\,H^2}  \frac{\partial G (\tau, \tau')}{\partial \tau} +   \frac{1}{\tau^2} \frac{V_{\phi\phi}(\phi_0)}{H^2} G (\tau, \tau')  =  \delta(\tau - \tau') \,,
\ee
with the boundary conditions $G (\tau',\tau' ) = 0$, $\partial G (\tau',\tau' )/\partial\tau =1$. The solution of Eq.~\eqref{greentext} is
\be\label{greent}
G (\tau, \tau') = \frac{\tau'}{\Delta}\left[ \left( \frac{\tau}{\tau'} \right)^{\nu_{+}} -\left( \frac{\tau}{\tau'} \right)^{\nu_{-}}\right]\Theta(\tau - \tau')\,,
\ee
where
\be \label{nuDelta}
\nu_{\pm} \equiv \frac{1}{2}\left(1-\frac{\pi\,\alpha\, V_{\phi}(\phi_0)}{f\,H^2} \right) \pm \frac{1}{2}\Delta\,, \quad \mbox{and} \quad \Delta \equiv \sqrt{\left(1-\frac{\pi\,\alpha\, V_{\phi}(\phi_0)}{f\,H^2} \right)^2 - \frac{4 V_{\phi\phi}(\phi_0)}{H^2}}\,.
\ee
Notice that, even if the homogeneous solution is suppressed by a $1/\Delta$ factor, the non-homogeneous solution sourced by Eq.~\eqref{delphi} is enhanced by the term $\frac{\alpha}{f}\delta_{\vec{E}\cdot\vec{B}}$, allowing perturbations can grow large.
Using Eq.~\eqref{greent} one can obtain the $n$-point correlator as
\ba \nonumber \label{ncorrelator}
\langle \delta \phi(\vec{p}_1)\cdots \delta \phi(\vec{p}_n) \rangle &=& \delta(\vec{p}_1 +\cdots + \vec{p}_n) \left(\frac{ {{\cal N}} \alpha }{f}\right)^n \int d\tau_1\cdots d\tau_n a_1^2 \cdots a_n^2 G(\tau, \tau_1) \cdots G(\tau, \tau_n) \\
& \times &\int d^3 x_1 \cdots  d^3 x_n e^{-i(\vec{p}_1\cdot\vec{x}_1 + \cdots + \vec{p}_n\cdot\vec{x}_n) }\langle \delta_{\vec{E}\cdot \vec{B} }(\tau_1, \vec{x}_1) \cdots \delta_{\vec{E}\cdot \vec{B} }(\tau_n, \vec{x})\rangle\,,
\ea
where $a_{i}\equiv a(\tau_i)$. 
\subsection{Power Spectrum} 
From Eq.~\eqref{ncorrelator} the spectrum of the perturbations is
\ba
\langle \delta \phi(\vec{p}) \delta \phi(\vec{p}\,') \rangle &=& \delta(\vec{p} + \vec{p}\,') \left(\frac{ {{\cal N}} \alpha }{f}\right)^2\int d\tau_1 d\tau_2 \,a_1^2\, a_2^2\, G(\tau, \tau_1)\,G(\tau, \tau_2)\nn\\
&&\qquad \times\int d^3x\, e^{i\vec{p}\cdot\vec{x}}\langle \delta_{\vec{E}\cdot \vec{B} }(\tau_1, 0)\, \delta_{\vec{E}\cdot \vec{B} }(\tau_2, \vec{x})\rangle\nonumber\\
& \approx & 
{\cal F(\nu_{+})}  \frac{\delta(\vec{p} + \vec{p}\,')}{p^3} \frac{ {{\cal N}^2}\, \alpha^2 \, H^4 }{   \Delta^2\, f^2\, \xi^8 }  e^{4\pi\, \xi} \, \left(-2^5\, \xi\, p\, \tau\right)^{2\nu_{+}} \,.
\ea
As discussed in Appendix~\ref{AA}, the dependence of $\mathcal{F}$ on $\nu_+$ is very small. We can thus safely take ${\cal F}\equiv{\cal F}(0)\approx2.13 \times 10^{-6}$.
With the previous results  we can calculate the power spectrum of the primordial curvature perturbation $\zeta = -H \,\delta\phi /\dot{\phi}_0$
\be
\delta(\vec{p} + \vec{p}\,'){\cal P}_{\zeta}(p) = \frac{p^3}{2 \pi^2} \frac{H^2}{\dot{\phi}_0^2} \langle \delta \phi(\vec{p})  \delta \phi(\vec{p}\,') \rangle \approx \delta(\vec{p} + \vec{p}\,')    \frac{{\cal F}}{2 \pi^2} \frac{H^2}{\dot{\phi}_0^2} \frac{ {{\cal N}^2} \alpha^2  H^4 }{   \Delta^2 f^2 \xi^8 }  e^{4\pi \xi} \left(-2^5 \xi p \tau\right)^{2\nu_{+}} .
\ee
Now, considering that $V_{\phi} \approx \frac{ {{\cal N}}\alpha}{f} \langle \vec{E}\cdot \vec{B} \rangle$ and using Eq.~\eqref{EBxi} one has
\be\label{V2xi}
\frac{  V^2_{\phi}(\phi_0) }{ {\cal I}^2 }\approx \frac{{\cal N}^2 \alpha^2 }{f^2} \frac{H^8}{\xi^8} \,e^{4\pi\xi}\,.
\ee
Expressing $\dot{\phi_0}$ in terms of $\xi$, and using  Eq.~\eqref{rootsapp} for $\Delta$ results in
\be\label{psscalet}
{\cal P}_{\zeta}(p) \approx    \frac{{\cal F} (\nu_{+})}{8 \pi^4\, {\cal I}^2 \, {\cal N}\, \xi^2}    \left(-2^5\, \xi\, p\, \tau\right)^{2\nu_{+}}  \approx  \frac{5\times 10^{-2}}{ {\cal N}\,\xi^2}   \left(-2^5\, \xi\, p\, \tau\right)^{2\nu_{+}},
\ee
where we used the fact that the contribution of the ${\cal N}$ gauge fields adds incoherently to the spectrum, which results in a suppression by a $1/{\cal N}$ factor. Here we notice that, as pointed out in Ref.~\cite{Anber:2009ua} for the minimally coupled case, in order to reproduce the observed amplitude of the scalar spectrum it is necessary to have a large number of gauge fields ${\cal N}\sim 10^5$, for $\xi\sim 10$. This is an unnatural feature of the model that persists in the presence of non-minimal coupling, as we will see in section~\ref{perturbations2}. Possible ways out of this problem are string theory inspired models containing a large number ${\cal N}$ of branes, each one with an associated $U(1)$ gauge field, or large symmetry groups like $SU(\sqrt{{\cal N}})$. Another possibility is to assume a slow-roll potential instead of the cosine-like, or to use auxiliary scalar fields instead of the inflaton coupled to the $U(1)$ gauge field, non-Abelian symmetry groups or massive vector fields~\cite{Anber:2009ua}.
Such scenarios are however not considered here.   

From the previous expression we extract the spectral index \cite{Anber:2009ua} \footnote{Notice that there is a difference of a global sign with respect to the result in Ref.~\cite{Anber:2009ua}.} 
\be\label{nsaphi}
n_s -1 \approx  { 2\nu_{+} } = {-}\frac{2 f\, V_{\phi\phi}(\phi_0)}{ \pi\,\alpha\, V_{\phi}(\phi_0) } \,.
\ee

\subsection{Tensor Perturbations} 
The calculation of the spectrum of the tensor perturbations in this model was originally done in Refs.~\cite{Sorbo:2011rz,Anber:2012du,Barnaby:2012xt}; here we briefly revisit the main details. The tensor perturbations with polarization $\lambda$ are obtained by solving the equation
\be\label{eqtenmod}
h''_\lambda-\frac{2}{\tau}h'_\lambda+k^2\, h_\lambda=\frac{2}{M^2_P}\Pi_\lambda^{lm}\,T_{lm}^\text{EM}, \quad \mbox{with} \quad h^{ij}(\vec{k})=\sqrt{2}\sum_{\lambda=\pm}\epsilon^i_\lambda(\vec{k})\,\epsilon^j_\lambda(\vec{k})\,h_\lambda(\tau,\vec{k})\,,
\ee 
where the energy momentum tensor ($T^\text{EM}$) and the projector ($\Pi_\lambda$) for the polarization plane are defined as
\be\nonumber
T_{ij}^\text{EM}= -a^2(E_i E_j + B_i B_j) + C\delta_{ij}=-a^{-2}A'_iA'_j+C\delta_{ij} \quad \mbox{and} \quad \Pi_\lambda^{lm}=\frac{1}{\sqrt{2}}\epsilon_{-\lambda}^l(\vec{k})\epsilon_{-\lambda}^m(\vec{k})\,.
\ee
The $C$-term does not contribute to the tensor perturbation because it is diagonal.
Eq.~\eqref{eqtenmod} can be rewritten as
\be\nonumber
 h_\lambda(\vec{k})=-\frac{2H^2}{M^2_P}\int d\tau' G_h(\tau,\tau')(-\tau')^{2}\int \frac{d^3q}{(2\pi)^{3/2}}\Pi_\lambda^{ij}A'_i(\vec{q},\tau')A'_j(\vec{k}-\vec{q},\tau')\;,
\ee
where $G_h(\tau,\tau')$ corresponds to the Green's function for the equation of the tensor perturbations~\eqref{eqtenmod} and is
\be\label{greentensor}
G_h(\tau,\tau') = \frac{1} {k^3\tau'^2} \Big[ (1+k^2\,\tau\, \tau')\,\sin\left[k(\tau -\tau')\right] + k(\tau -\tau')\,\cos\left[k(\tau -\tau') \right]\Big]\Theta(\tau,\tau')\,,
\ee
with $\Theta$ being the Heaviside's function. 
Using Eq.~\eqref{Asol} we get the tensor perturbation spectrum \cite{Sorbo:2011rz}
\be
{\cal P}^{t\pm} = \frac{H^2}{\pi^2 M_P^2}\left( 1 + {\cal A}^{\pm}  \frac{ {\cal N} H^2}{M_P^2}\frac{e^{4\pi \xi}}{\xi^6}\right)\,,
\ee
where ${\cal A}^{+}\approx 8.6\times10^{-7}$ and ${\cal A}^{-}\approx 1.8\times10^{-9}$ are constants for each mode polarization. They are different as a consequence of the parity violating nature of this model. With them, we obtain the tensor-to-scalar ratio 
\be
r = \frac{{\cal P}^{t+} + {\cal P}^{t-} }{{\cal P}_{\zeta}} =\frac{H^2}{\pi^2 M_P^2} \frac{2+ ({\cal A}^{+} + {\cal A}^{-}) \frac{ {\cal N} H^2}{M_P^2}\frac{e^{4\pi \xi}}{\xi^6} }{{\cal P}_{\zeta}} \approx  \frac{H^2}{\pi^2 M_P^2} \frac{2+ {\cal A}^{+}  \frac{ {\cal N} H^2}{M_P^2}\frac{e^{4\pi \xi}}{\xi^6} }{{\cal P}_{\zeta}}\,.
\ee 
Using the approximate Friedmann equation $H^2 \approx V/(3M_P^2)$ together with Eqs.~\eqref{V2xi} and~\eqref{psscalet} we obtain \cite{Anber:2012du}:
\be\label{rVV1}
r  \approx  \frac{2 V}{3\pi^2 M_P^4\, {\cal P}_{\zeta}} + \frac{72 \pi^2\, {\cal A}^{+}}{{\cal F}} \frac{\xi^4}{\alpha^2} \left( \frac{f \,V_{\phi}}{V} \right)^2  \approx  \frac{2 V}{3\pi^2 M_P^4\, {\cal P}_{\zeta}} +2.9\times 10^2 \frac{\xi^4}{\alpha^2} \left( \frac{f\, V_{\phi}}{V} \right)^2 , 
\ee
where $\mathcal{P}_\zeta\simeq \frac{0.05}{\mathcal{N}\,\xi^2}\simeq 2.5\times 10^{-9}$ is the amplitude of the power spectrum with the so called COBE normalization. This comes   from Eq.~\eqref{psscalet}, neglecting the scale dependent part.  In Eq.~\eqref{rVV1} we can notice the separation of the vacuum gravitational waves spectrum and the gravitational waves sourced by the coupling with the gauge fields. The sourced part is amplified for large values of $\xi$ and have a definite polarization: they are chiral. Let us emphasize that $\xi$ does not affect the spectrum of the vacuum gravitational waves, but only the sourced component.  
 
\subsection{Numerics of the Minimally Coupled Case} 
In this section we inspect the parameter space of this model. We use the spectral index $n_s$ in Eq.~\eqref{nsaphi} and the tensor-to-scalar ratio $r$ in Eq.~\eqref{rVV1} to compare with current Planck bounds~\cite{Ade:2015lrj}.
To this end, we approximate the value of $\xi$ by solving the equation of motion of the inflaton~\eqref{phitau} neglecting time derivatives
\begin{equation}
\xi\simeq\frac{1}{2\pi}\log\left[\frac{1.3\times 10^4}{{\cal N}\,\alpha}\,\xi^4\,\frac{M_P^2}{H^2}\frac{f\, |V_{\phi}|}{V}\right]\,,
\end{equation}
and use it to compute the slow-roll parameters
\ba
\epsilon &\equiv& -\frac{\dot{H}}{H^2}=  \frac{  2  \xi^2\, f^2  }{ \alpha^2\, M_{P}^2  } + \frac{8}{7}\frac{\xi}{\alpha}\frac{f\, V_{\phi}}{V}\,, \\ \nonumber
\eta&\equiv&-\frac{\ddot\phi}{\dot\phi\,H} = {\epsilon} - \frac{1}{2\,  {H}\, {\epsilon}} \frac{d {\epsilon}}{d {t}}=-\frac{1}{H}\left[\frac{\dot H}{H}+\frac{\dot\xi}{\xi}\right] \\
& \simeq& -\frac{1}{H}\left[\frac{f \,\xi}{\sqrt{3}\,\alpha \,M_P}\frac{V_{\phi}}{\sqrt{V}}+\frac{-2\frac{\dot H}{H}V_{\phi}+\dot\phi\left(V_{\phi\phi}-\frac{V_{\phi}^2}{V}\right)}{2\pi\,M_P^3-\frac{4}{\xi}V_{\phi}}\right]\,.
\ea
The field excursion can be related with the number $N_e$ of $e$-folds that lasts inflation 
\begin{equation}
N_e\simeq\int_{\phi_i}^{\phi_f}\frac{H}{\dot\phi}d\phi=\frac{\alpha}{2\,f}\int_{\phi_i}^{\phi_f}\frac{1}{\xi}d\phi\simeq\frac{\alpha}{2\,\xi}\frac{\phi_f-\phi_i}{f}\,,
\end{equation}
given that the rate $H/\dot{\phi}$ is approximately constant. 

Hereafter we consider the natural inflation potential~\cite{Freese:1990rb, Adams:1992bn}
\be\label{potnatinfl}
V(\phi) = {\Lambda^4} \left[1+ \cos\left(\frac{\phi}{f}\right)\right],
\ee
since, with this choice, it is possible to generate slow-roll due to the friction caused by the axial coupling term. This implies that, differently from the case of the original natural inflation, one can have a steep potential, which means that $f$ can acquire sub-Planckian values. 

\begin{figure}[h!]
	\centering
	\includegraphics[width=0.47\textwidth]{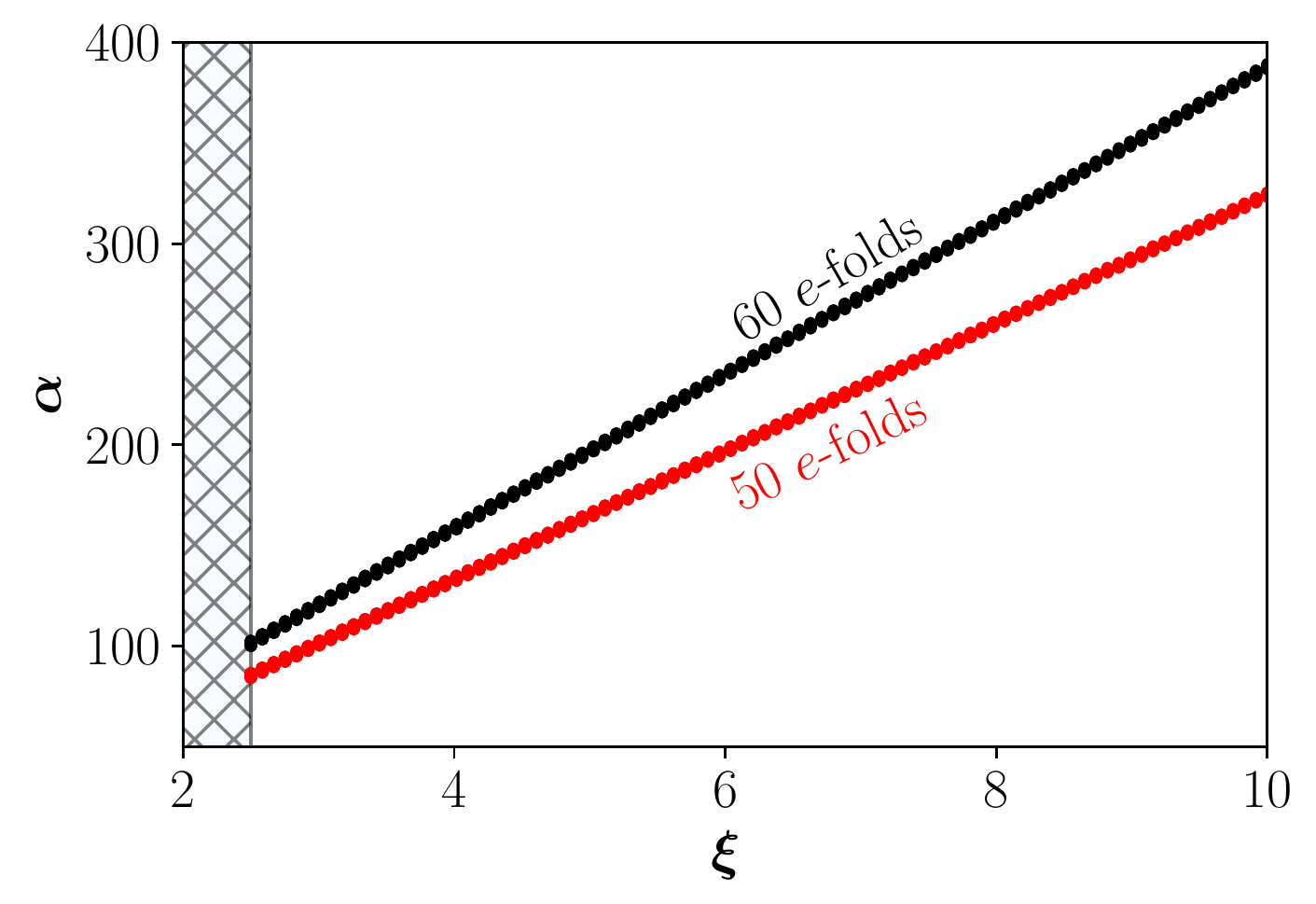}\\
	\includegraphics[width=0.47\textwidth]{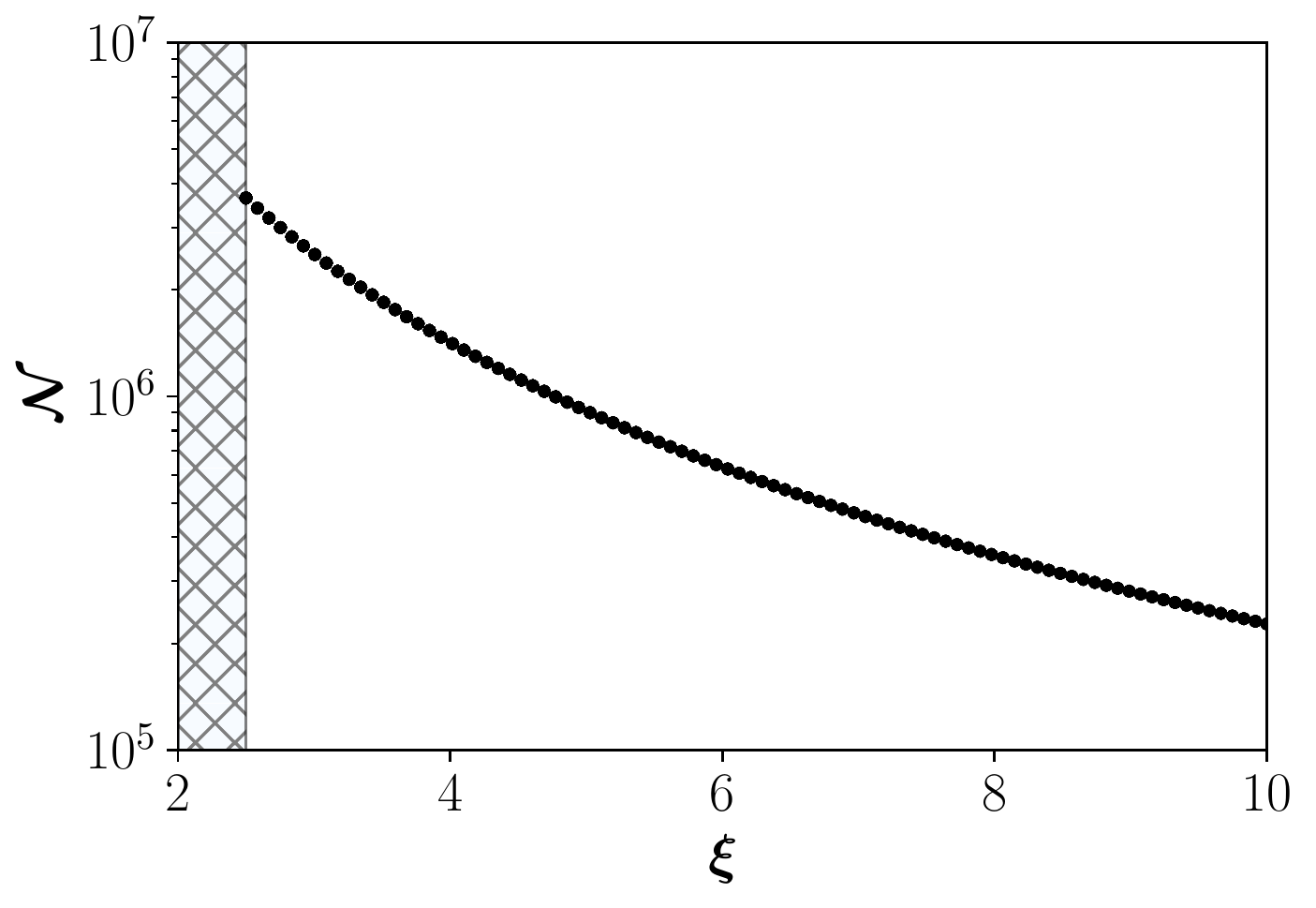}
	\includegraphics[width=0.47\textwidth]{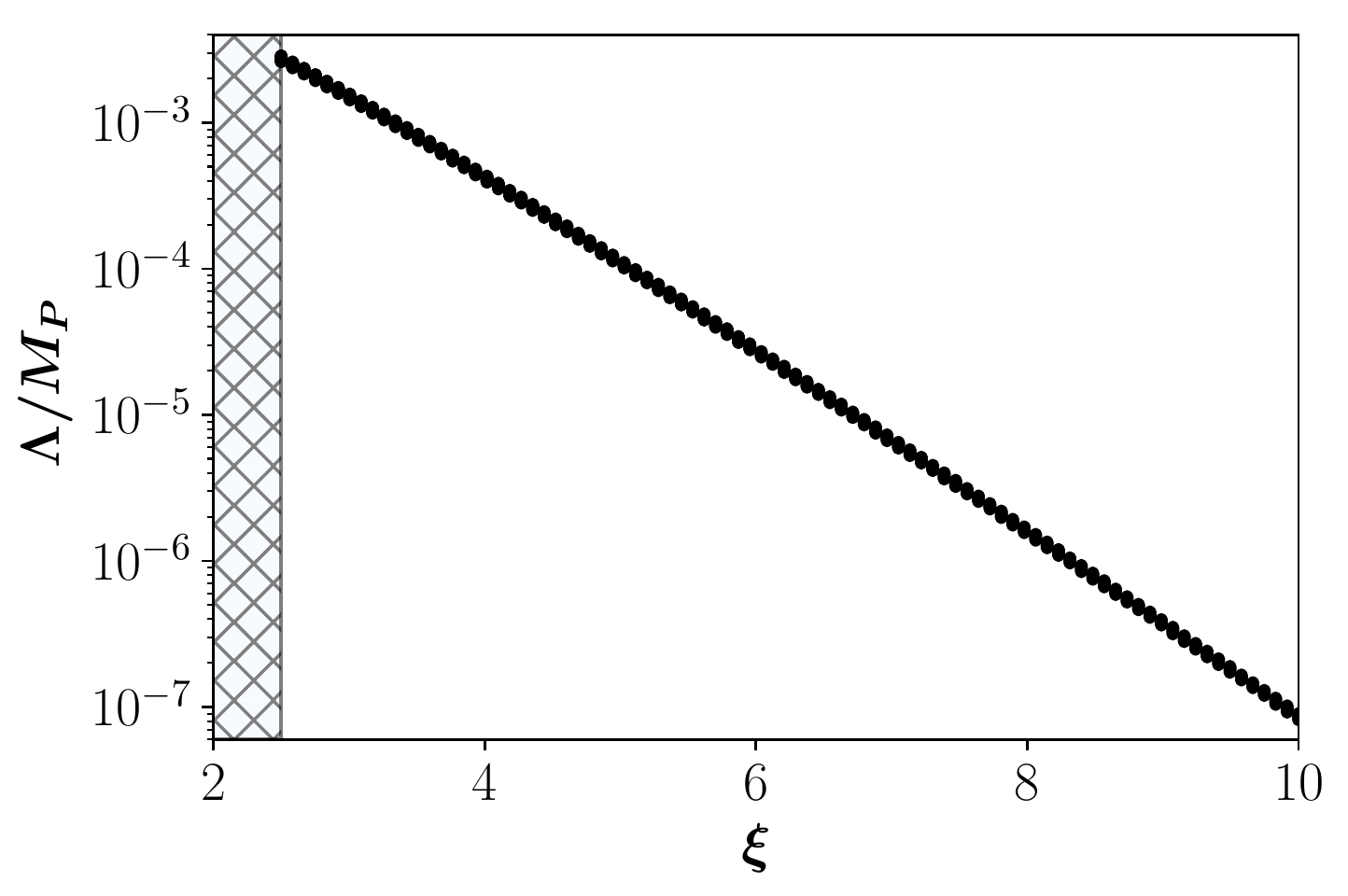}
	\caption{Values of $\xi$, $\alpha$, $\mathcal{N}$ and $\Lambda$ compatible with Planck limits at $68\%$ CL.
	Note that while $\mathcal{N}$ is independent from the number of $e$-folds, $\Lambda$ is almost insensitive to that parameter.
	The region $\xi<2.5$ is beyond the validity of the current approximations and hence disregarded. Large $\xi$ are considered since the slow-roll evolution is driven by the gauge field production, obtained precisely for large $\xi$ values.
	}
	\label{fig:xi_photon}
\end{figure}

For the sake of completeness, it is instructive to do a more precise statement about the regime in which the backreaction term dominates the slow-roll evolution. To this end, we compare the derivative of the potential term against the backreaction term in the RHS of Eq.~\eqref{phitau}. Using the cosine potential and the estimate for the backreaction term in Eq.~\eqref{EBxi} we find that the friction term dominates when
\begin{equation}
\frac{e^{2\pi \xi}}{\xi^4} > \frac{9 M_{P}^4 }{ {\cal N I} \alpha\Lambda^4 }.
\end{equation}
As an example, even if we take a single gauge field ${\cal N}=1$, $\alpha = 200$ and $\Lambda=4.5\times 10^{-3} M_{P}$ (which correspond to typical values, see e.g. Figs.~\ref{fig:xi_photon} and~\ref{fig:xi_0and10}), we get that the friction term dominates when $\xi\gtrsim5.4$.

\begin{figure}[t!]
	\centering
	\includegraphics[width=0.47\textwidth]{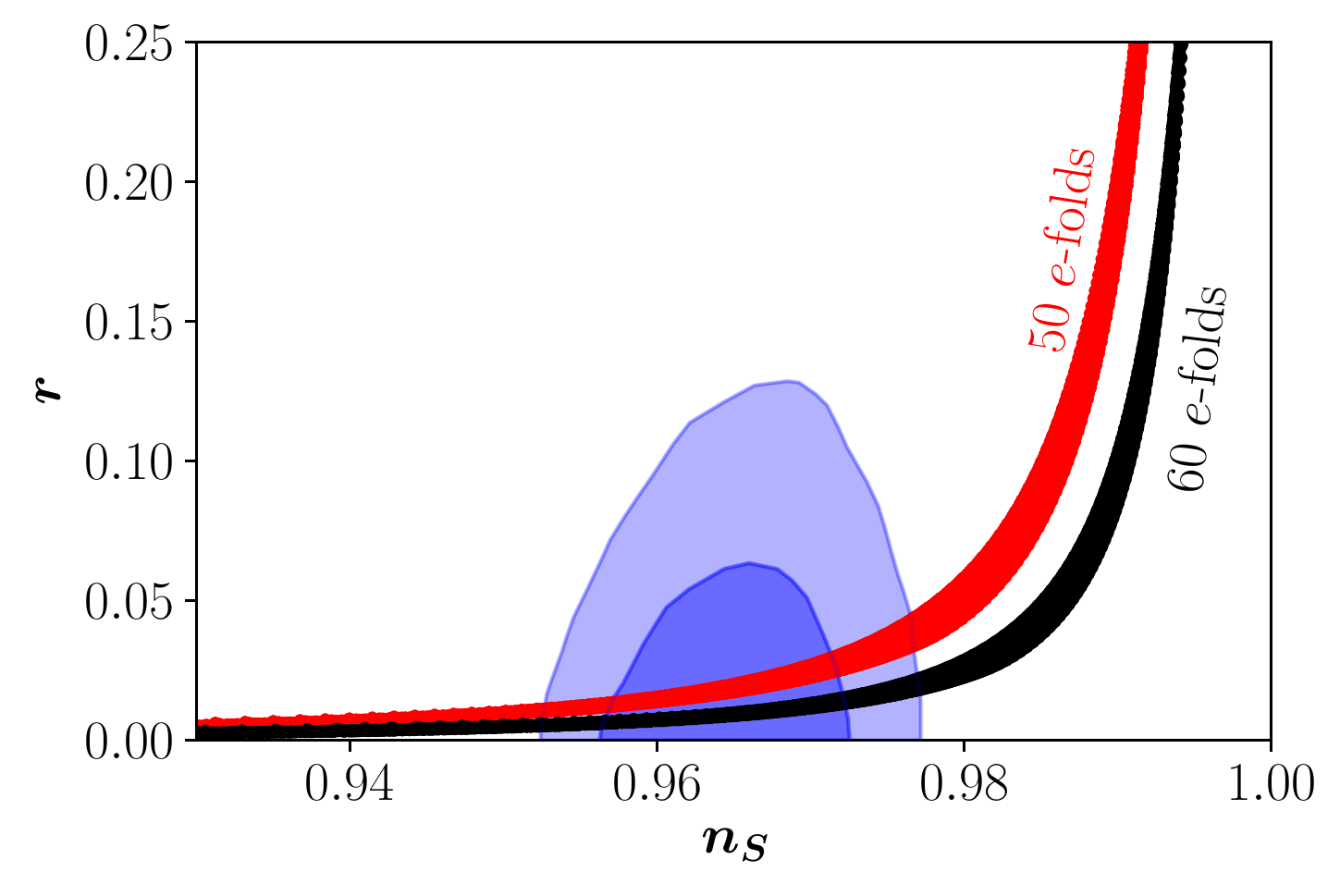}
	\caption{Tensor-to-scalar ratio $r$ versus the scalar spectral index $n_s$ for 50 and 60 $e$-folds, in the case of natural inflation with the inflaton coupled to $\mathcal{N}$ $U(1)$ gauge fields.
	The blue bands correspond to the $68\%$ (light) and $95\%$ (dark) CL regions from Planck.
	}
	\label{fig:r-ns_photon}
\end{figure}

Fig.~\ref{fig:r-ns_photon} shows the $68\%$ (light blue) and $95\%$ (dark blue) CL regions for the tensor-to-scalar ratio $r$ versus the scalar spectral index $n_s$ from Planck~\cite{Ade:2015lrj}.
The predictions in the case of natural inflation with the inflaton coupled with $\mathcal{N}$ gauge fields are also shown, for 50 (red band) and 60 $e$-folds (black band).
These regions were constructed by varying the parameters in the ranges: $\xi=[2.5,\,10]$ and $\alpha=[80,\,400]$.
For a given number of $e$-folds, the number $\cal N$ of gauge fields is fixed by the COBE normalization, the scale $\Lambda$ of the potential by the equation of motion of the inflaton.
Let us also note that in the minimally coupled case, both $r$ and $n_s$ are independent on the coupling $f$ of the potential, in the regime where $\xi$ is almost constant.
It is clear from the figure that there are regions compatible with the $95\%$ CL Planck limits, for the cases with both 50 and 60 $e$-folds.\footnote{For this to be possible, the minus sign in the expression for the spectral index in Eq. (\ref{nsaphi}) is necessary, otherwise, there is no region in the parameter space compatible with Planck's constraints.}

Moreover, Fig.~\ref{fig:xi_photon} depicts the values of $\xi$, $\alpha$, $\mathcal{N}$ and $\Lambda$ compatibles with Planck limits at $68\%$ CL.
Note that while $\mathcal{N}$ is independent from the number of $e$-folds, $\Lambda$ is almost insensitive to that parameter.
The region $\xi<2.5$ is beyond the validity of the current approximations and hence not considered.
Let us point out that the number of $U(1)$ gauge fields required to reproduce the amplitude of the scalar spectrum has to be about $\mathcal{N}\sim 10^6$.
The inflationary scale $\Lambda$ is always sub-Planckian, going down to $\mathcal{O}(10^{-7})\,M_P$ for $\xi\simeq 10$.
Additionally, the coupling constant $\alpha$ is of the order $10^2$; this is however not a problem per se for perturbativity, because it always appears suppressed by $f$, which is at the Planck scale.

\subsection{Comments on Non-Gaussianities} 
For the sake of completeness, we present the non-Gaussianities produced by this model~\cite{Anber:2012du, Barnaby:2010vf, Barnaby:2011vw}.
Using Eq.~\eqref{ncorrelator} we calculate the 3-point function of the scalar perturbations. In the equilateral limit
\be
\langle \delta \phi(\vec{p}_1)\, \delta \phi(\vec{p}_2)\, \delta \phi(\vec{p}_3) \rangle^{\rm (equil)} \propto {\cal F}_{3} \frac{\delta(\vec{p}_1 + \vec{p}_2 +  \vec{p}_3)}{p^6} \frac{  \alpha^3\,  H^6\, e^{6\pi \xi} }{    f^3 \,\xi^{9} }    \,,
\ee
where ${\cal F}_{3}\approx 5\times 10^{-11}$ \cite{Barnaby:2010vf, Barnaby:2011vw}. The bispectrum of the primordial curvature perturbation is then
\be
\langle \zeta (\vec{p}_1)\, \zeta(\vec{p}_2)\, \zeta (\vec{p}_3) \rangle^{\rm (equil)} \propto - {\cal F}_{3} \frac{H^3}{\dot{\phi_0}^3} \frac{\delta(\vec{p}_1 + \vec{p}_2 +  \vec{p}_3)}{p^6} \frac{  \alpha^3 \, H^6\, e^{6\pi \xi} }{    f^3\, \xi^{9} }    \,.
\ee

For an arbitrary potential able to support slow-roll the $f_{\rm NL}$ parameter in the equilateral configuration is given by~\cite{Barnaby:2010vf, Barnaby:2011vw}
\be
f_{\rm NL}^{\rm equil} \approx 4.4\times 10^{10} \,{\cal P}^3 \,\frac{e^{6\pi \xi}}{\xi^9}\,,
\ee
where ${\cal P}^{1/2} = H^2/(2\pi\dot{\phi}_0)$. The current bound on equilateral non-Gaussianity is $f_{\rm NL}^{\rm equil} = -4\pm 43$ \cite{Ade:2015ava}.
This is a stringent constraint for this kind of models since it requires $\xi \lesssim 2.5$. 
For such small values of the parameter $\xi$, the signatures from the chiral tensor perturbations sourced by the coupling with the axial term $\phi \tilde{F}F$ are suppressed and indistinguishable from the vacuum perturbations.

However, for models strongly coupled to gauge fields where the slow-roll is achieved due to the friction term, the equilateral non-Gaussianity is given by~\cite{Anber:2012du}
\be
f_{\rm NL}^{\rm equil} \approx -1.3\, \xi\,, 
\ee 
which allows us to consider values above $\xi \approx 2.5$, where $\xi$ reaches a plateau and can be fairly enough approximated by a constant. 
This is the case considered here.
In the next section, the situation in the non-minimal coupling case is discussed.

\section{Including Non-minimal Coupling with Gravity} 
\label{nmcg}
We consider the inclusion of a non-minimal coupling between the pseudoscalar field and gravity.
The Lagrangian for the gravity-scalar system is
\ba\label{lsgc} 
{\cal L}= \sqrt{-g}\left[  \frac{{M_P^{2}}}{2}\left( 1  + \frac{2 h(\phi)}{M_P^2} \right){R} - \frac{1}{2}(\partial \phi)^2 - V(\phi)  -\frac{\mathcal{N}}{4} F^{\mu \nu}F_{\mu \nu} -\frac{\mathcal{N}\,\alpha }{4f}\phi F^{\mu \nu}\tilde{F}_{\mu \nu}  \right]\,,
\ea
where $h(\phi)$ introduces a particular form of non-minimal coupling with gravity. In the following we restrict to an $h(\phi)$ of the form 
\ba\label{nmch} 
h(\phi) =  \frac{1}{2} \nm\, \phi^2   \,,
\ea
where $\nm$ is a dimensionless constant, the non-minimal coupling parameter between gravity and the pseudoscalar field.
This leads to the equation of motion for the pseudoscalar field
\be\label{eomsc} 
\Box \phi -   V_{\phi} +  \nm R\, \phi -\frac{\mathcal{N}\,\alpha }{4f} F^{\mu \nu}\tilde{F}_{\mu \nu}    =  0\,.
\ee
Let us note that for a free field (i.e. $V_{\phi}= 0$ and $\alpha =0$), 
this equation is invariant under conformal transformations when $\nm = -1/6$. Moreover, the form of the gauge field equation remains unaltered by the non-minimal coupling to gravity 
\be\label{aeomnm}
\frac{ d^2 A_i}{d {\tau}^2} - {\nabla}^2 A_i - \frac{\alpha }{f }\frac{d \phi}{d {\tau}}  ({\nabla}\times A)_{i} = 0\,.
\ee
However, the effect of the non-minimal coupling is introduced only through the dynamics of the pseudoscalar field, i.e. through the coupling parameter $\xi$ which depends both on the time and on the non-minimal coupling parameter. 

\subsection{Jordan and Einstein Frames} \label{jtoe}
The action (\ref{lsgc}) describes the system in the so-called Jordan frame in which the non-minimal coupling between the pseudoscalar field and gravity appears explicitly. We can do a conformal scaling of the metric in the following form: 
\be
\bar{g}_{\mu\nu} = \Omega(\phi) \, g_{\mu\nu}, \quad \mbox{where} \quad \Omega(\phi) \equiv 1+ \frac{2 h(\phi)}{M_P^{2}} \,.
\ee
For non-minimal couplings of the form (\ref{nmch}), the conformal factor is
\be
\Omega(\phi) = 1 + \nm \left( \frac{ \phi}{M_P} \right)^2 \,.
\ee
Under this transformation the action (\ref{lsgc}) becomes 
\be\label{eomef}
{\cal L} = \sqrt{-\bar{g}} \left[  \frac{{M_P^{2}}}{2}    \bar{R}  -  \frac{1}{2}K(\phi)\, \bar{g}^{\mu \nu}\, \bar{\nabla}_{\mu} \phi\, \bar{\nabla}_{\nu} \phi -  \bar{V}(\phi)  - \frac{\mathcal{N}}{4 } \bar{F}^{\mu \nu}\bar{F}_{\mu \nu} - \frac{\mathcal{N}\,\alpha }{4f } \phi    \bar{F}^{\mu \nu}\tilde{\bar{F}}_{\mu \nu}  \right], 
\ee
where $\bar{V} (\phi) \equiv \frac{V(\phi)}{\Omega^2}$, $\bar{F}^{\mu \nu}\bar{F}_{\mu \nu} \equiv \bar{g}^{\mu \alpha}\bar{g}^{\nu\beta} F_{\alpha \beta}F_{\mu\nu}$ and $\bar{F}^{\mu \nu}\tilde{\bar{F}}_{\mu \nu} \equiv \bar{g}^{\mu \alpha}\bar{g}^{\nu\beta} F_{\alpha \beta}\tilde{F}_{\mu\nu}$. The details of the derivation of Eq.~\eqref{eomef} are discussed in Appendix \ref{AB}. For a single pseudoscalar we can write the Lagrangian as a canonical pseudoscalar field coupled minimally to gravity and to the vector field through the axial term 
\be\label{eomefcan}
{\cal L} = \sqrt{-\bar{g}} \left[  \frac{{M_P^{2}}}{2}    \bar{R}  -  \frac{1}{2}\bar{g}^{\mu \nu}\, \bar{\nabla}_{\mu} \bar{\phi}\, \bar{\nabla}_{\nu} \bar{\phi} -  \bar{V}(\bar{\phi})  - \frac{\mathcal{N}}{4 } \bar{F}^{\mu \nu}\bar{F}_{\mu \nu} - \frac{\mathcal{N}\,\alpha}{4f } \, \phi(\bar{\phi})  \,  \bar{F}^{\mu \nu}\tilde{\bar{F}}_{\mu \nu}  \right], 
\ee
where the canonical pseudoscalar field $\bar{\phi}$ is defined through the transformation $d \bar{\phi} /d \phi  = K ^{1/2}$. The action (\ref{eomefcan}) is now in the Einstein frame~\cite{Kaiser:2010ps}: it is written using only fields with canonical kinetic terms and a metric that allows the action to acquire the traditional form of the Einstein-Hilbert gravity. 
 Using the particular form of the non-minimal coupling function~\eqref{nmch}, we get the transformation function
\be\label{Kfunction}
 K(\phi)  =  \frac{1}{\Omega} + \frac{6 }{ M_P^2\, \Omega^2}  ( \nm\, \phi)^2 = \frac{  1+6( \nm + \frac{1}{6}) \, \nm \left(\frac{ \phi}{M_P}\right)^2   }{ \left[ 1 +\nm \left(\frac{ \phi}{M_P}\right)^2 \right]^2} \,.
\ee
\subsection{Equations of Motion in the Einstein Frame} 
We can now derive the equations of motion for the canonical fields in the Einstein frame from the action (\ref{eomefcan}). 
It is important to derive the equation of motion for the canonical field $\bar{\phi}$ since, for this field, the construction of the perturbation correlations are obtained following the usual canonical quantization procedure~\cite{Makino:1991sg}. Since the action is written in terms of the coordinates $\bar{x}$ and the metric $\bar{g}_{\mu \nu}$, all the derivatives are compatible with this metric.  The result for the pseudoscalar field is:
\be
\bar{\Box} \bar{\phi} -   \bar{V}_{ \bar{\phi}} - \frac{{\cal N}\, \alpha }{4 f\, {K^{1/2} }} \bar{F}^{\mu \nu}\tilde{\bar{F}}_{\mu \nu} = 0\,,
\ee
where $\bar{\Box} \bar{\phi} \equiv \bar{\nabla}^{\mu} \bar{\nabla}_{\mu} \bar{\phi} =\frac{1}{\sqrt{-\bar{g}}} \,\bar{\partial}_{\mu} (\sqrt{-\bar{g}} \bar{g}^{\mu \nu}\, \bar{\partial}_{\nu} \bar{\phi})$. 
Now, we consider the isotropic solution
\be \label{barFL}
 d \bar{s}^2 = - d \bar{t}\,^2 + \bar{a}^2 \delta_{ij} d \bar{x}_i  d\bar{x}_j =   \bar{a}^2(\bar{\tau})\left[- d \bar{\tau}^2 +  \delta_{ij} d \bar{x}_i  d\bar{x}_j \right].
\ee 
To go from the Jordan to the Einstein frame variables for the isotropic solution (\ref{barFL}) we take into account that $d \bar{s}^2 = \Omega \,d {s}^2 $, which implies
\be\label{attauOm}
 d \bar{t} = \Omega^{1/2}\, d t , \quad \bar{a}(\bar{t}) = \Omega^{1/2}\, a(t), \quad  d \bar{\tau} =   d \tau \quad \mbox{and}\quad  \bar{H} = \frac{1}{\bar{a}} \frac{ d \bar{a}}{ d \bar{t}} = \frac{1}{\Omega^{3/2}} \left[  H \,\Omega +  \frac12\dot{\Omega}  \right]\,.
\ee 
In conformal coordinates, the equation for the canonical pseudoscalar field becomes
\be\label{nmccphi}
\frac{ {\partial^2 \bar{\phi}} }{\partial \bar{\tau}^2}  + 2\bar{a}\,\bar{H} \,\frac{ {\partial \bar{\phi}} }{\partial \bar{\tau}} - \bar{\nabla}^2 \bar{\phi}  +  { \bar{a}^2 }\, \bar{V}_{ \bar{\phi}}   =  \frac{ \bar{a}^2 }{ {\Omega^2 \,K^{1/2}} }\frac{ {\cal N} \alpha }{ f } \vec{E}\cdot \vec{B}\,,
\ee 
where we use the definition for the electric and magnetic field components as in Eqs.~\eqref{EBdef} and~\eqref{EBid} in the Jordan frame. 
For gravity, we derive the Einstein equations varying with respect to $\bar{g}_{\mu\nu}  $:
\be
\bar{R}_{\mu\nu} - \frac{1}{2} \bar{g}_{\mu\nu} \bar{R} = \frac{1}{M_P^2} \bar{T}_{\mu\nu}\,,
\ee
with the energy momentum tensor
\be\label{bemtensor}
\bar{T}_{\mu\nu} \equiv \frac{-2 }{\sqrt{-\bar{g}}} \frac{ \delta {\cal L} }{\delta \bar{g}^{\mu\nu}} =    \bar{\partial}_{\mu} \bar{\phi}\, \bar{\partial}_{\nu} \bar{\phi} - \bar{g}_{\mu\nu}\left( \frac{1}{2}\bar{g}^{\alpha \beta} \bar{\partial}_{\alpha} \bar{\phi}\, \bar{\partial}_{\beta} \bar{\phi} + \bar{V}  \right) + \bar{g}^{\alpha \beta} {F}_{\mu\alpha} {F}_{\nu\beta}  - \bar{g}_{\mu\nu} \frac{1}{4 } \bar{F}^{2}\,,
\ee
which leads to the Friedmann equations for the metric $\bar{g}_{\mu\nu}$ 
\ba \label{febar1}
\bar{H}^2 &=& \frac{1}{3 M_P^2} \left[ \frac{1}{2}  \left(\frac{\partial \bar{\phi}}{\partial \bar{t}} \right)^2  + \bar{V} + \frac{1}{2{\Omega^2}}\langle \vec{E}^2 + \vec{B}^2 \rangle \right]\,,   \\ \label{febar2}
\frac{d {\bar{H}} }{d\bar{t}}   &=& - \frac{1}{  2 M_P^2}   \left(\frac{\partial  \bar{\phi} }{\partial \bar{t}} \right)^2   - \frac{1}{ 3 M_P^2 \,{\Omega^2}}   \langle \vec{E}^2 + \vec{B}^2 \rangle\,.
\ea
Finally, for the equation of motion for the gauge fields it is important to notice that  the derivatives with respect to the conformal time and with respect to spacial coordinates in both frames are the same, then the equation obtained is the same Eq.~\eqref{aeomnm}. If one goes to Fourier space and projects into the transverse polarizations, one gets the same Eq.~\eqref{apm} obtained in the minimally coupled case
\be 
A''_{\pm}  + \left( k^2  \pm \frac{2 k\, \xi}{\tau} \right)A_{\pm} = 0, \quad \mbox{with} \quad \xi \equiv \frac{\alpha \,\dot{\phi_0}}{2f\, H}\,.
\ee
Notice that in this equation appears the velocity of the pseudoscalar field $\dot{\phi_0}$ and not the velocity of the canonical field. Both quantities can be related through the expression
\be\label{xitobxi}
\xi = \frac{\alpha\, \dot{\phi_0}}{2f\, H} \approx \frac{\alpha }{2f } \frac{1}{\bar{H}\, K^{1/2}} \frac{d \bar{\phi}_0}{d\bar{t} } \equiv \frac{1}{ K^{1/2} } \bar{\xi}\,. 
\ee

\subsection{Perturbations} \label{perturbations2}
Starting from Eq.~\eqref{nmccphi}, we write the equations for the background canonical field and its perturbation $\bar{\phi}(\bar{x}, \bar{\tau}) = \bar{\phi}_0(\bar{\tau}) + \delta\bar{\phi}(\bar{x}, \bar{\tau})$
\ba \label{nmphi0}
\bar{\phi}''_0 + 2\bar{a}\,\bar{H}\,\bar{\phi}'_0   +  \bar{a}^2\, \bar{V}_{ \bar{\phi}} & = &   \frac{ \bar{a}^2 }{ {\Omega^2 \,K^{1/2}} } \frac{ {{\cal N}}\alpha}{f} \langle \vec{E}\cdot \vec{B} \rangle\,. \\ \label{nmphipert}
  \delta \bar{\phi}'' + 2\bar{a}\, \bar{H} \,\delta\bar{\phi}'  +  \bar{a}^2\, \bar{V}_{ \bar{\phi}\bar{\phi}}\, \delta \bar{\phi}  & = & \frac{ \bar{a}^2\, {\cal N}\,\alpha }{ f} \left[ \delta\left[ \frac{ 1 }{ {\Omega^2  }  K^{1/2}} \right] \langle \vec{E}\cdot \vec{B} \rangle + \frac{ 1 }{ {\Omega^2  }  K^{1/2}} \delta\left[\vec{E}\cdot \vec{B} \right]     \right],\qquad 
\ea
where the primes here represent derivatives with respect to the conformal time in the Einstein frame $\bar{\tau}$. We neglected the gradient term since we assume that the perturbations are homogeneous. 
As we did for the minimally coupled case, we calculate de dependence of the term $\vec{E}\cdot \vec{B} $ with the velocity of the canonical field
\be
\delta\left[\vec{E}\cdot \vec{B} \right] \approx   \delta_{\vec{E}\cdot \vec{B} }+ \frac{\partial  \langle \vec{E}\cdot \vec{B} \rangle }{\partial (\partial \bar{\phi}/ \partial \bar{t} ) }   \frac{\partial  \delta \bar{\phi} }{ \partial \bar{t} }\,.
\ee
The second term can be approximated as 
\be
\frac{\partial  \langle \vec{E}\cdot \vec{B} \rangle }{\partial (\partial \bar{\phi}/ \partial \bar{t} ) }   \frac{\partial  \delta \bar{\phi} }{ \partial \bar{t} } \approx  \frac{\partial  \langle \vec{E}\cdot \vec{B} \rangle }{\partial \xi }  \frac{\partial  \xi }{\partial (\partial \bar{\phi}/ \partial \bar{t} ) } \frac{ \delta \bar{\phi}' }{ \bar{a} } \approx    2\pi \langle \vec{E}\cdot \vec{B} \rangle \frac{\partial  \xi }{\partial (\partial \bar{\phi}/ \partial \bar{t} ) } \frac{ \delta \bar{\phi}' }{ \bar{a} } \approx  \frac{\alpha }{f } \frac{\pi \langle \vec{E}\cdot \vec{B} \rangle}{ K^{1/2}} \frac{ \delta \bar{\phi}' }{ \bar{H} \bar{a} }  \,,
\ee
where we have used Eq.~\eqref{xitobxi} in the last step. 
We can relate the potential with the value of the term $\langle \vec{E}\cdot \vec{B} \rangle$ through the equation for the background field $\bar{\phi}_0$. Neglecting time derivatives in Eq.~\eqref{nmphi0} and using the approximation $ \bar{V}_{ \bar{\phi}} \approx   \frac{ {{\cal N}}\alpha}{  {\Omega^2 K^{1/2}} f} \langle \vec{E}\cdot \vec{B} \rangle $, the perturbation of the source  term becomes
\be
\delta\left[\vec{E}\cdot \vec{B} \right] \approx   \delta_{\vec{E}\cdot \vec{B} }+\pi \frac{\Omega^2}{{\cal N}}\, \bar{V}_{ \bar{\phi}} \, \frac{ \delta \bar{\phi}' }{ \bar{H}\, \bar{a} }\,.
\ee
With these results, the equation for the perturbations can be rewritten as 
\be 
 \label{nmphipertap}
  \delta \bar{\phi}'' + 2\bar{a} \,\bar{H}\left( 1-\frac{\pi \,\alpha\, \bar{V}_{ \bar{\phi}} }{ 2K^{1/2} \,f\, \bar{H}^2}  \right)  \delta\bar{\phi}'  +  \bar{a}^2 \left(  \bar{V}_{ \bar{\phi} \bar{\phi} }  + \frac{d \ln ( {\Omega^2}K^{1/2} )}{ d \bar{\phi}}   \bar{V}_{ \bar{\phi}} \right) \delta \bar{\phi}  = \frac{ \bar{a}^2\, {\cal N}\,\alpha }{ {\Omega^2\,  K^{1/2} } \,f} \delta_{\vec{E}\cdot \vec{B} }\,,
\ee
where it is understood that all the terms depending of the canonical pseudoscalar field are evaluated at $\bar{\phi}_0$. 
The formal solution to this equation is
\be
\delta \bar{\phi}(\bar{\tau}, \vec{k}) =  \frac{ {\cal N} \alpha}{f} \int_{-\infty}^{\bar{\tau}} d\bar{\tau}_1 \frac{\bar{a}^2(\bar{\tau}_1)\, \bar{G}(\bar{\tau}, \bar{\tau}_1)}{ \Omega^2 \,K^{1/2} } \int d^3x\, e^{-i\vec{k}\cdot\vec{x}} \, \delta_{\vec{E}\cdot \vec{B} }(\bar{\tau}_1, \vec{x})\,, 
\ee
where the Green function is obtained from
 \be 
 \label{nmgreen}
  \left[ \frac{d^2 }{d \bar{\tau}^2} -  \frac{2}{\bar{\tau}}\left( 1-\frac{\pi \alpha \bar{V}_{ \bar{\phi}} }{ 2K^{1/2} f \bar{H}^2}  \right)     \frac{d }{d \bar{\tau}} +  \frac{ \bar{V}_{ \bar{\phi} \bar{\phi}} }{\bar{H}^2\bar{\tau}^2 } \left(  1  +  \frac{d \ln ( {\Omega^2}K^{1/2} )}{ d \bar{\phi}} \frac{ \bar{V}_{ \bar{\phi}} }{ \bar{V}_{ \bar{\phi} \bar{\phi}} } \right) \right]  \bar{G}(\bar{\tau}, \bar{\tau}')  = \delta(\bar{\tau} - \bar{\tau}')\,.
\ee
The solution to this expression with the boundary conditions $\bar{G}(\bar{\tau}', \bar{\tau}' )=0$, $\bar{G}'(\bar{\tau}', \bar{\tau}' )=1$ is 
\ba \label{nmgreensol}
\bar{G}(\bar{\tau}, \bar{\tau}' ) &=& \frac{\bar{\tau}'}{ \bar{\Delta} }\left[ \left( \frac{ \bar{\tau} }{ \bar{\tau}' } \right)^{\bar{\nu}_{+}} -\left( \frac{ \bar{\tau} }{ \bar{\tau}' } \right)^{\bar{\nu}_{-}}\right]\Theta( \bar{\tau} - \bar{\tau}'), \quad \mbox{with}\\
 \nu_{\pm} &\equiv& \frac{1}{2}\left(1-\frac{\pi\,\alpha\,  \bar{V}_{ \bar{\phi}}( \bar{\phi}_0)}{ K^{1/2} f\, \bar{H}^2 } \right) \pm \frac{1}{2}\bar{\Delta}\approx -\frac{1}{2}\left(\frac{\pi\,\alpha\,\bar{V}_{ \bar{\phi}}( \bar{\phi}_0)}{ K^{1/2} f \bar{H}^2 } \right) \pm \frac{1}{2}\bar{\Delta}\,,\\ \nonumber
 \bar{\Delta} &\equiv& \sqrt{\left(1-\frac{\pi\,\alpha\,\bar{V}_{ \bar{\phi}}(\bar{\phi}_0)}{ K^{1/2} f\,\bar{H}^2 } \right)^2 - \frac{4}{ \bar{H}^2} \left(   \bar{V}_{ \bar{\phi}\bar{\phi}} ( \bar{\phi}_0) +\frac{d \ln (  {\Omega^2}K^{1/2} )}{ d \bar{\phi}}  \bar{V}_{ \bar{\phi}} (\bar{\phi}_0)\right) } \\ \label{Deltanm}
&\approx & \sqrt{\left(\frac{\pi\,\alpha\,  \bar{V}_{ \bar{\phi}}( \bar{\phi}_0)}{ K^{1/2} f \bar{H}^2 } \right)^2 - \frac{4}{ \bar{H}^2} \left(   \bar{V}_{ \bar{\phi}\bar{\phi}} ( \bar{\phi}_0) +\frac{d \ln (  {\Omega^2}K^{1/2} )}{ d \bar{\phi}}  \bar{V}_{ \bar{\phi}} (\bar{\phi}_0)\right) }\,,
\ea
where we used the approximation $\frac{\pi\, \alpha\, \bar{V}_{ \bar{\phi}} }{ 2K^{1/2} f\, \bar{H}^2}  \gg 1$. The late time evolution of the Green's function is dominated by the $\nu_{+}$ term, so we can approximate the Green's function with
\be
\bar{G}(\bar{\tau}, \bar{\tau}' ) \approx \frac{\bar{\tau}'}{ \bar{\Delta} } \left( \frac{ \bar{\tau} }{ \bar{\tau}' } \right)^{\bar{\nu}_{+}} \quad \mbox{for} \qquad \bar{\tau}>\bar{\tau}'\,.
\ee
Following the same steps of the minimally coupled case, we calculate the spectrum of the scalar perturbations
\ba 
\langle \delta \bar{\phi}(\vec{p}) \delta \bar{\phi}(\vec{p}\,') \rangle  \approx  
 {\cal F(\bar{\nu}_{+})}  \frac{\delta(\vec{p} + \vec{p}\,')}{p^3} \frac{ {\cal N}^2\, \alpha^2\,  \bar{H}^4 }{  K\, \bar{\Delta}^2 f^2\, \xi^8 } \, e^{4\pi \xi}\,  \left(-2^5 \xi \,p\, \bar{\tau}\right)^{2\bar{\nu}_{+}}.
\ea
At this point one can calculate the spectrum of the primordial curvature perturbation $\bar{\zeta} = - \bar{H}\, \delta \bar{\phi}/\dot{\bar{\phi}}_0$.
However, it has been shown  that it is unaltered by a conformal transformation~\cite{Makino:1991sg}, meaning that this variable and its correlators are the same in Einstein and Jordan frames:\footnote{See e.g. Ref.~\cite{Kamenshchik:2014waa} for a discussion of quantum equivalence between Jordan frame and Einstein frame.}
\be
{\cal \bar{P}}_{\zeta}(p)  \approx     \frac{{\cal F}}{2 \pi^2} \frac{\bar{H}^2}{\dot{\bar{\phi}}_0^2} \frac{ {{\cal N}^2} \alpha^2  \bar{H}^4 }{  K \bar{\Delta}^2 f^2 \xi^8 }  e^{4\pi \xi} { \left(-2^5 \xi\, p\, \bar{\tau}\right)^{2 \bar{\nu}_{+}}},
\ee
which using Eqs.~\eqref{EBxi},~\eqref{Deltanm} and  $ \bar{V}_{ \bar{\phi}} \approx   \frac{ {{\cal N}}\alpha}{  {\Omega^2 K^{1/2}} f} \langle \vec{E}\cdot \vec{B} \rangle $ gives
\be\label{psscalenm}
{\cal \bar{P} }_{\zeta}(p) \approx    \frac{{\cal F} ( \bar{\nu}_{+})  }{8 \pi^4 {\cal I}^2 {\cal N}\, \xi^2}   { \left(-2^5 \xi\, p\, \tau\right)^{2  \bar{\nu}_{+} }} \approx  \frac{5\times 10^{-2}  }{ {\cal N} \xi^2}  { \left(-2^5 \xi\, p\, \tau\right)^{2  \bar{\nu}_{+}}} . 
\ee
The spectral index can then be extracted:
\be\label{nsnmc}
\bar{n}_s -1 \approx  { 2\bar{\nu}_{+} } ={-} K^{1/2} \frac{2 f\, \bar{V}_{ \bar{\phi}\bar{\phi}}(\bar{\phi}_0)}{ \pi\,\alpha\,  \bar{V}_{ \bar{\phi}}(\bar{\phi}_0) } \left( 1 + \frac{d \ln ( {\Omega^2}K^{1/2} )}{ d \bar{\phi}} \frac{  \bar{V}_{ \bar{\phi}} (\bar{\phi}_0) }{ \bar{V}_{ \bar{\phi}\bar{\phi}}(\bar{\phi}_0)}\right) . 
\ee
We can see from the previous results that the amplitude of the scalar perturbation is kept unaltered:
\be
 {\cal \bar{P} }_{\zeta} \approx  
 {\cal {P} }_{\zeta}\,,
\ee
but the spectral index is modified by a $K^{1/2}$ factor. It is important to notice that, even though the expression for the spectrum is equal both for minimal and for non-minimal couplings, the dynamics of the  parameter $\xi$ is different in the two cases. 
\begin{figure}[t!]
	\centering
	\includegraphics[width=0.47\textwidth]{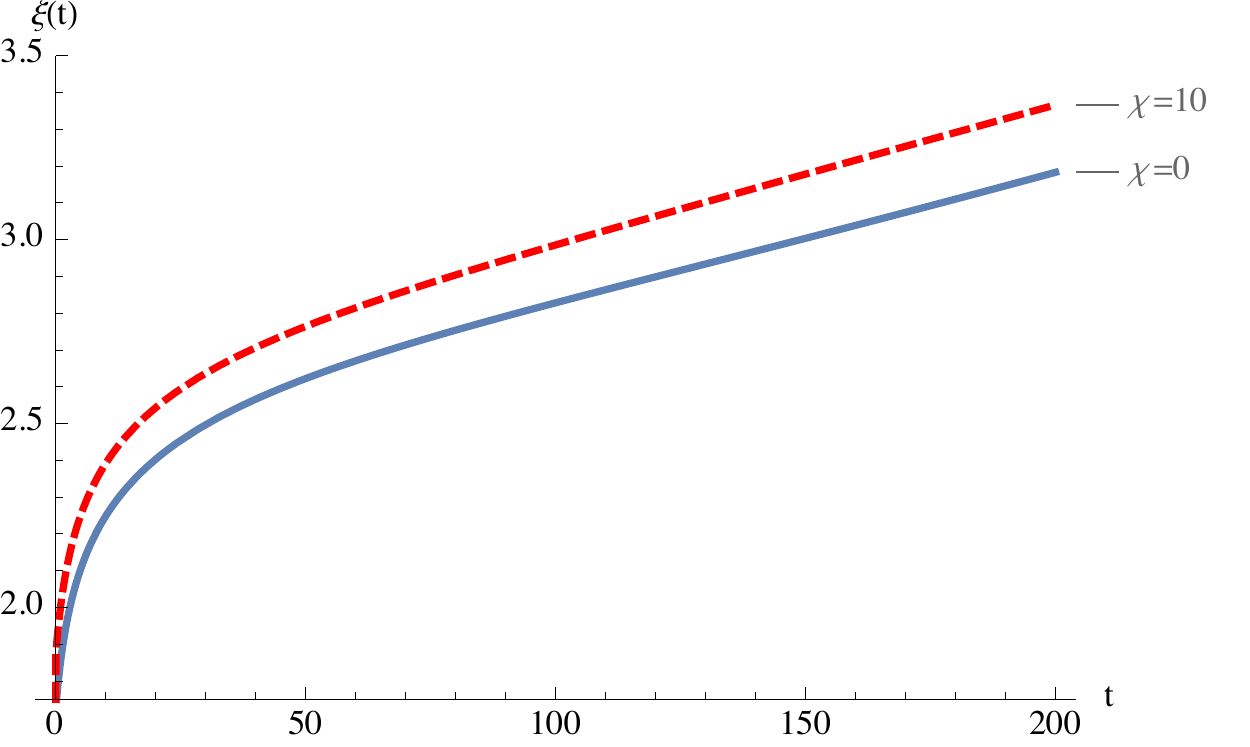}
	\caption{An example of the evolution of the parameter $\xi$ for minimal coupling (solid blue) and non-minimal coupling with $\nm = 10$ (dashed red). We use the values $\Lambda=4.5\times 10^{-3}M_{P}$, $\alpha=400$, ${\cal N}=10^5$, $f=0.1\,M_{P}$. The time is measured in units of $M_{P}/\Lambda^2$.
	}
	\label{fig:xi_0and10}
\end{figure}
Fig.~\ref{fig:xi_0and10} reflects that difference with an example of the evolution of $\xi$ with $\nm=0$ (solid blue line) and $\nm=10$ (dotted red line). These solutions are obtained by solving Eq.~\eqref{eomsc}, with $H$ obtained from the Friedman Eq.~\eqref{febar1} and using the approximations~\eqref{Asol} and~\eqref{EBxi}. We also choose $\Lambda=4.5\times 10^{-3}M_{P}$, $\alpha=400$, ${\cal N}=10^5$ and $f=0.1\, M_{P}$. As it can be seen, even for a large value of the non-minimal coupling such as $\nm=10$, the difference is not very significant, in the ballpark of few percents. Note that throughout the paper we used the assumption of a constant $\xi$. Fig.~\ref{fig:xi_0and10} shows that this assumption for late times is justified, nevertheless, a more precise statement of this assumption can be made if we consider the ratio $\dot{\xi}/(H\xi)$. This analysis is shown in the Appendix~\ref{constantxi}.

\subsection{Tensor Perturbations} 
In this section we discuss how non-minimal couplings can modify the production of sourced gravitational waves. Knowing that the tensor perturbation variable $h_{\lambda}$ is not altered by conformal transformations, we can do the calculation for the metric
\be \label{hnm}
 d \bar{s}^2 = -  \bar{a}^2(\bar{\tau})\left[- d \bar{\tau}^2 + ( \delta_{ij} + \bar{h}_{ij})  d \bar{x}_i  d\bar{x}_j \right].
 \ee
As in the minimally coupled case, the spectrum is obtained by solving the Eq.~\eqref{eqtenmod} with the metric (\ref{hnm}) and with the energy momentum tensor for the sourced part as in Eq.~\eqref{bemtensor}:
\be 
\bar{T}^\text{EM}_{ij}= -\bar{a}^2 ( E_i E_j + B_i B_j ) + C \delta_{ij} = -\frac{1}{\bar{a}^2} A'_{i}A'_{j} + C \delta_{ij}\,.
\ee
The result is the same that the one obtained in the minimally coupled case but, in the conformal coordinates ($\bar{\tau}, \bar{x}_i$), it is:
\be
\bar{{\cal P}}^{t\pm} = \frac{\bar{H}^2}{\pi^2 M_P^2}\left( 1 + {\cal A}^{\pm}  \frac{ {\cal N} \bar{H}^2}{M_P^2}\frac{e^{4\pi \xi}}{\xi^6}\right),
\ee
then, the total tensor spectrum reads
\be
\bar{{\cal P}}^{t} \approx \frac{2\bar{H}^2}{\pi^2 M_P^2} \left({1+\frac{({\cal A}^{+} + {\cal A}^{-})}{2}  \frac{ {\cal N} \bar{H}^2}{ M_P^2}\frac{e^{4\pi \xi}}{\xi^6} }\right)\,,
\ee
and the tensor-to-scalar ratio
\be\label{barr}
\bar{r} = \frac{\bar{{\cal P}}^{t+} + \bar{{\cal P}}^{t-} }{\bar{{\cal P}}_{\zeta}} =\frac{2\bar{H}^2}{\pi^2 M_P^2} \frac{1+ \frac{({\cal A}^{+} + {\cal A}^{-})}{2} \frac{ {\cal N} \bar{H}^2}{ M_P^2}\frac{e^{4\pi \xi}}{\xi^6} }{\bar{{\cal P}}_{\zeta}} \approx  \frac{2\bar{H}^2}{\pi^2 M_P^2} \frac{1+ \frac{{\cal A}^{+}}{2}  \frac{ {\cal N} \bar{H}^2}{ M_P^2}\frac{e^{4\pi \xi}}{\xi^6} }{\bar{{\cal P}}_{\zeta}}\,.
\ee 
We can express the previous results in terms of the potential by using the approximate Friedmann equation $\bar{H}^2 \approx \bar{V}/(3M_P^2)$, the scalar spectrum  $\bar{{\cal P}}_{\zeta} \approx    {\cal P }_{\zeta}$ of Eq.~\eqref{psscalenm}
and the derivative of the potential 
\be
 \bar{V}_{ \bar{\phi}} \approx   \frac{ {{\cal N}}\alpha}{ f\, {\Omega^2\, K^{1/2}} } \langle \vec{E}\cdot \vec{B} \rangle \approx - \frac{ {\cal N} \alpha\, {\cal I} }{ f\, {\Omega^2 \,K^{1/2}} } \frac{H^4}{\xi^4} e^{2\pi \xi}\,.
\ee  
With them, we can write
\be\label{rVV1nmc} 
\bar{r}   \approx  \frac{2 \bar{V}}{3\pi^2 M_P^4  \,{\cal P}_{\zeta}} + \frac{72 \pi^2 {\cal A}^{+}\, {K} }{{\cal F}} \frac{\xi^4}{\alpha^2} \left( \frac{f \,\bar{V}_{\bar{\phi}}}{\bar{V}} \right)^2  \approx    \frac{2 {V}}{3\pi^2 \,M_P^4 \, \Omega^2 \,{\cal P}_{\zeta}} + 2.9\times 10^2  {K} \frac{\xi^4}{\alpha^2} \left( \frac{f\, \bar{V}_{\bar{\phi}}}{\bar{V}} \right)^2.
\ee
It is worth emphasizing that the first term of the last equation, the one corresponding to the vacuum fluctuations, is suppressed by the factor $\Omega^2$ in the denominator.
The second term, the one responsible for the `sourced' tensor perturbations, is modified by the factor $K$. This factor changes significantly for small field values depending on the value of the non-minimal coupling parameter $\nm$. This difference in the behavior between the vacuum and sourced terms offers a possibility for generate observable chiral sourced gravitational waves which would acquire an enhancement due to the non-minimal coupling. We can look for a region in the parameters space in which the current observational constrains over $n_s$ and $r$ are respected. We explore this possibility in the next section.

\section{Numerical Analysis of the Non-minimal Coupled Case}
\label{numnmc}

\subsection{Natural Inflation} 
Before going to the full system with the axial coupling, it is interesting to consider a single pseudoscalar field driving inflation. We do not introduce any other auxiliary field as source of the primordial curvature perturbation or any spectator field.  The most widely studied example of small field inflation with a pseudo-Nambu-Goldstone scalar, able to produce an inflationary expansion is natural inflation \cite{Freese:1990rb, Adams:1992bn}, characterized by a potential given in Eq.~\eqref{potnatinfl}.
We also introduce a coupling with gravity of the form $h(\phi)R=\frac12 \nm \,\phi^2\, R$.%
\footnote{This coupling is still consistent with parity breaking, but breaks the shift symmetry.
A related approach that preserves the tree-level shift symmetry is discussed in Refs.~\cite{Germani:2010hd,Folkerts:2013tua}. Instead of introducing the non-minimal coupling in the form of a function $h(\phi)R$, they use a non-minimal derivative coupling of the type $K(\phi) \partial \phi \partial \phi$, keeping the scale of the coupling constant $f$ far below from Planck scale and achieving in that way an UV protected completion of the theory. Other possibilities also include coupling to Gauss-Bonnet and Chern-Simons terms.}

We are interested in the observables $n_s$ and $r$ related with the spectrum of the scalar and the tensor perturbations respectively. To this end, we need to evaluate the slow-roll parameters for this potential:
\begin{eqnarray}
\bar{\epsilon}_V &\equiv& \frac{M_P^2}{2 K}\left(\frac{\bar V_\phi}{\bar V}\right)^2
=\frac{\left[4 f\,\nm\,\phi+(\nm\,\phi^2+M_P^2)\,\tan\frac{\phi}{2 f}\right]^2}{2 f^2\,\left[M_P^2+\nm\,\phi^2\,(6\nm+1)\right]}\,, \\
\bar{\eta}_V&\equiv&\frac{M_P^2}{K\,\bar V}\left(\bar V_{\phi\phi}-\frac{K'}{2 K}\bar V_\phi\right)\nonumber\\
&=&\frac{1}{2M_P\,f\left(M_P^2+\nm\,\phi^2(6\nm+1)\right)^2}\nonumber\\
&&\times\Bigg[2f\,\nm\,\phi(\nm\,\phi^2+M_P^2)(7M_P^2+6M_P^2\,\nm+7\nm\,\phi^2(6\nm+1))\tan\frac{\phi}{2f}\nonumber\\
&&\qquad-8f^2\,\nm(M_P^4-\nm\,\phi^2\,(3M_P^2+4\nm\,\phi^2(6\nm+1)))\nonumber\\
&&\qquad+(\nm\,\phi^2+M_P^2)^2(M_P^2+\nm\,\phi^2(6\nm+1))\sec^2\frac{\phi}{2f}\nonumber\\
&&\qquad-2(\nm\,\phi^2+M_P^2)^2(M_P^2+\nm\,\phi^2(6\nm+1))\Bigg]\,.
\end{eqnarray}
We use the fact that the slow-roll parameters associated with the potential $\epsilon_V$ and $\eta_V$ are approximately equal to the Hubble parameter slow-roll $\epsilon_H\approx \epsilon_V$ and $\eta_H \approx \eta_V - \epsilon_V$. Moreover, we rely here on the facts that the primordial curvature perturbation observables are invariant under a conformal transformation~\cite{Makino:1991sg} and that the slow-roll variables in the canonical variables of the Einstein frame are invariant as well~\cite{Chiba:2008ia}: $\epsilon_V=\bar\epsilon_V$ and $\eta_V \approx \bar{\eta}_V$.

We express our results in terms of the number of $e$-folds of the inflationary expansion:
\be
N_e=-\frac{1}{M_P^2}\int_{\phi_i}^{\phi_f}K\,\frac{\bar V}{\bar V_\phi}\,d\phi\,,
\ee
and we let the field roll from the moment at which the perturbations crosses the horizon at $k=a\,H$, until the moment in which the slow-roll condition are broken: $\bar{\epsilon}_{V}(\phi)>1$ or $\bar{\eta}_{V}(\phi) >1$.
The spectral index of the scalar perturbations $n_s$ and the tensor-to-scalar ratio $r$ are obtained form the slow-roll parameters as
\be
\bar{n}_s = n_s=1-6\bar{\epsilon}_V+2\bar{\eta}_V \quad \mbox{and}\quad \bar{r} = r =16\,\bar{\epsilon}_V\,.
\ee

\begin{figure}[t!]
	\centering
	\includegraphics[width=0.47\textwidth]{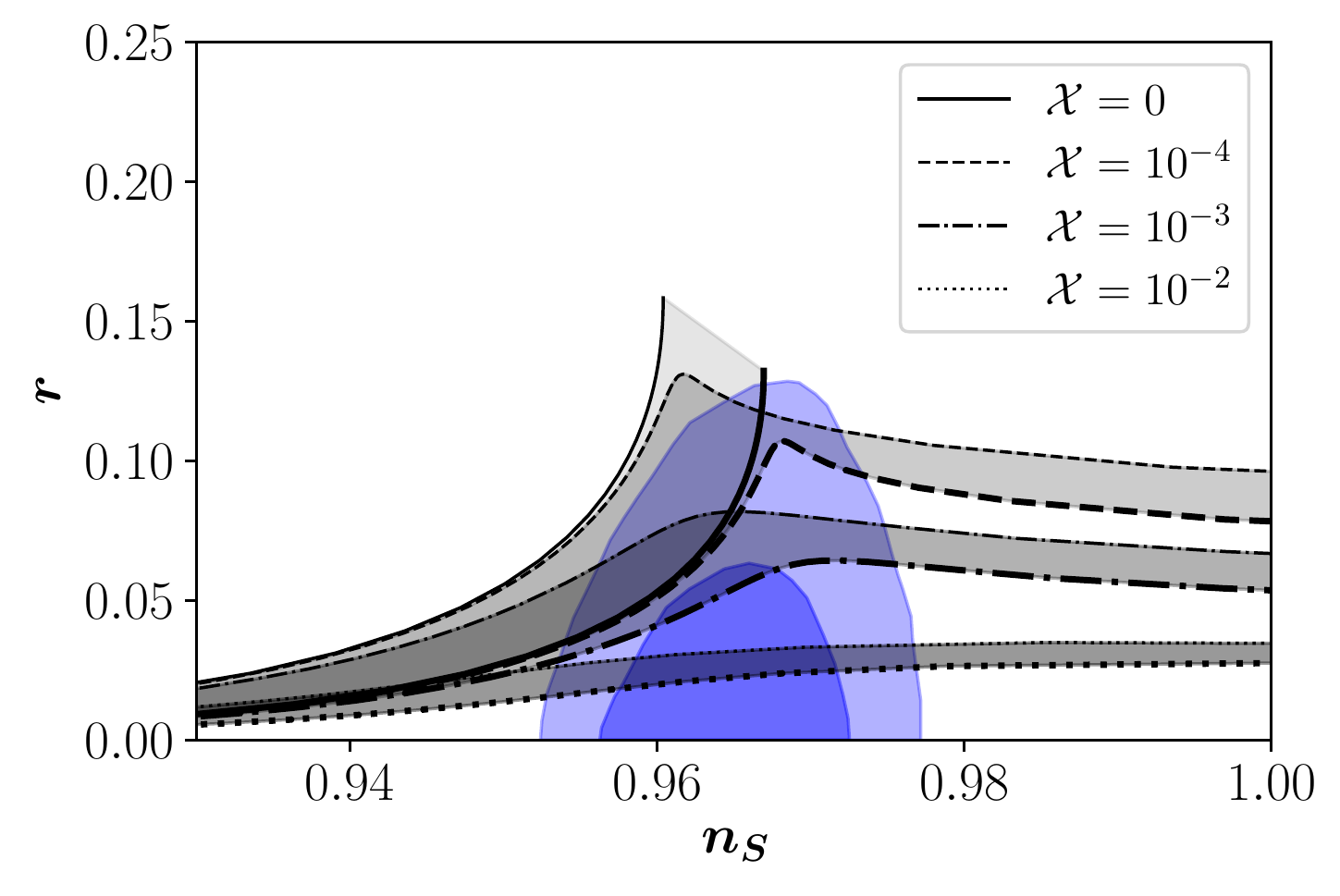}
	\includegraphics[width=0.47\textwidth]{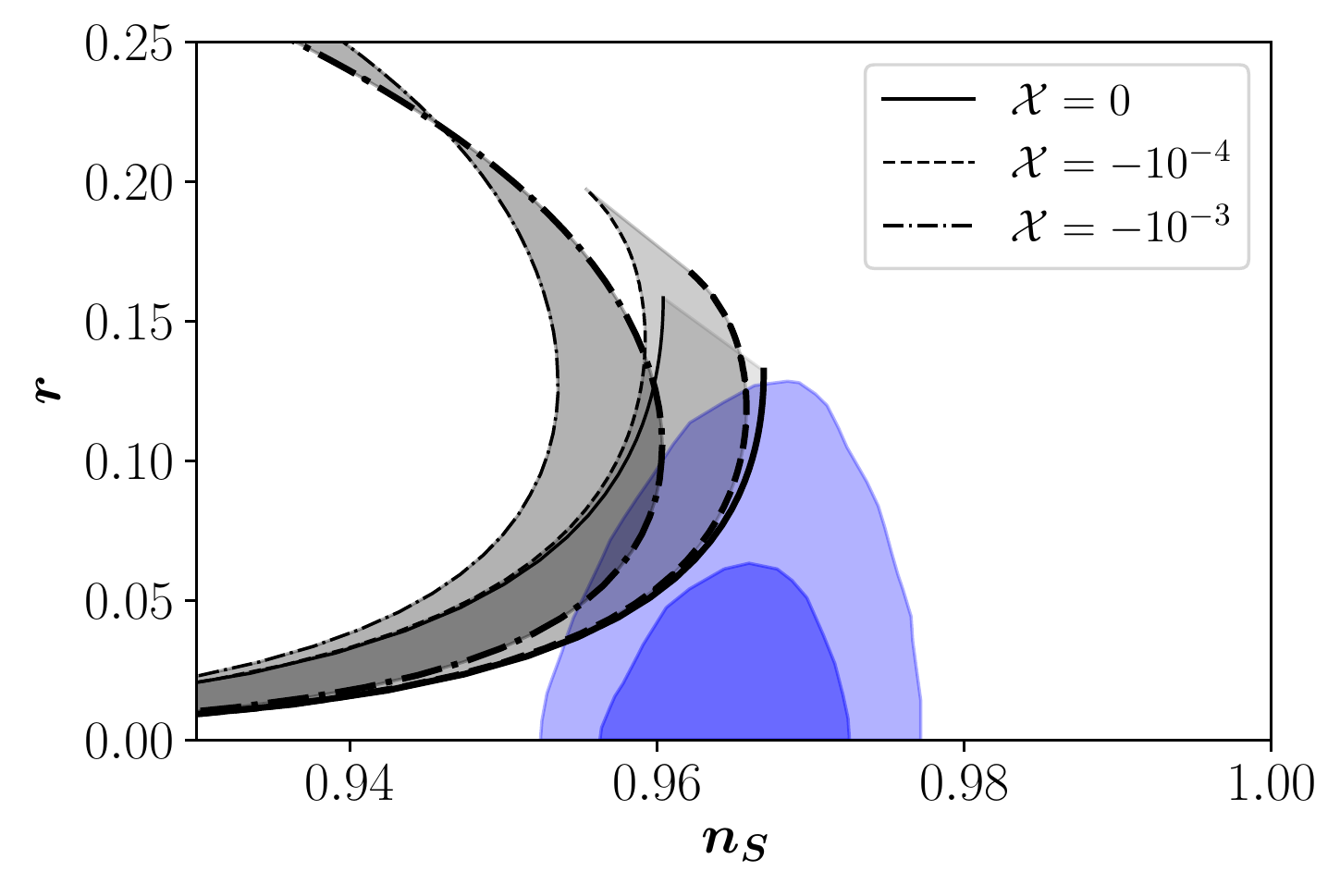}
	\caption{Tensor-to-scalar ratio $r$ versus the scalar spectral index $n_s$, in the case of natural inflation with non-minimal coupling $\nm$.
	Black thin and thick lines correspond to 50 and 60 $e$-folds, respectively.
	The blue bands correspond to the $68\%$ (light) and $95\%$ (dark) CL regions from Planck.
	}
	\label{fig:r-ns}
\end{figure}

Fig.~\ref{fig:r-ns} shows the $68\%$ (light blue) and $95\%$ (dark blue) CL regions for the tensor-to-scalar ratio $r$ versus the scalar spectral index $n_s$ from Planck.
The predictions in the case of natural inflation with non-minimal coupling are also shown, for different values of $\nm$.
The back thin and thick lines correspond to 50 and 60 $e$-folds, respectively.\footnote{An analysis of natural inflation with non-minimal coupling was done before in Ref.~\cite{Nozari:2015ada}. Even if we agree with their analytical results, we diverge on the numerics.}
We use small values for $\nm$ since in this small field case, the dynamics is sensible to small values of the non-minimal coupling.
On one hand, we see that positive values of $\nm$ tend to suppress the amplitude of tensor perturbations and to render the scalar perturbations closer to the scale invariance.
On the other hand, negative values of $\nm$ tend to enhance the production of tensor mode perturbations and to diverge from scale invariance. The case of negative $\nm$ is disfavored by current Planck constraints.

\begin{figure}[t!]
	\centering
	\includegraphics[width=0.47\textwidth]{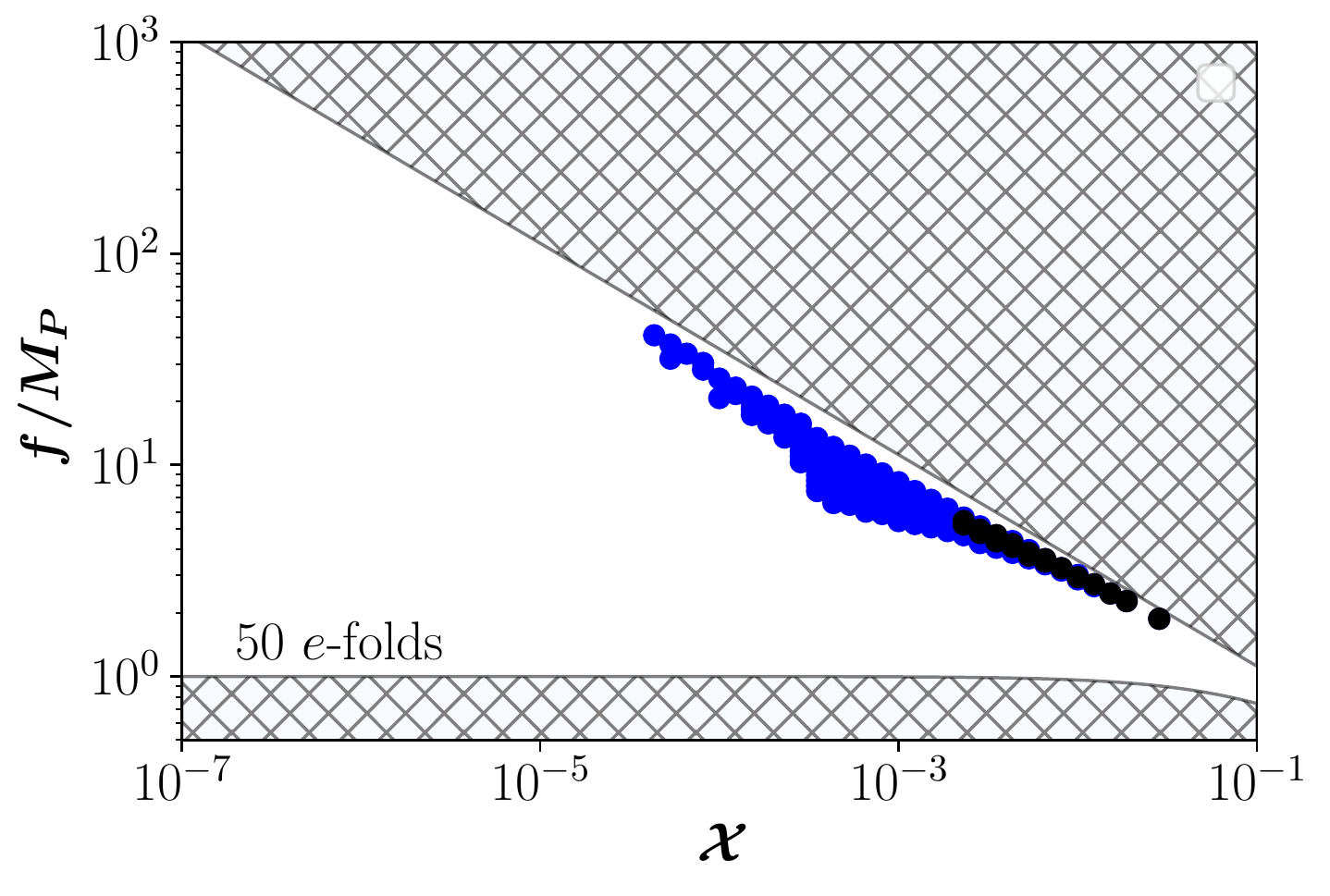}
	\includegraphics[width=0.47\textwidth]{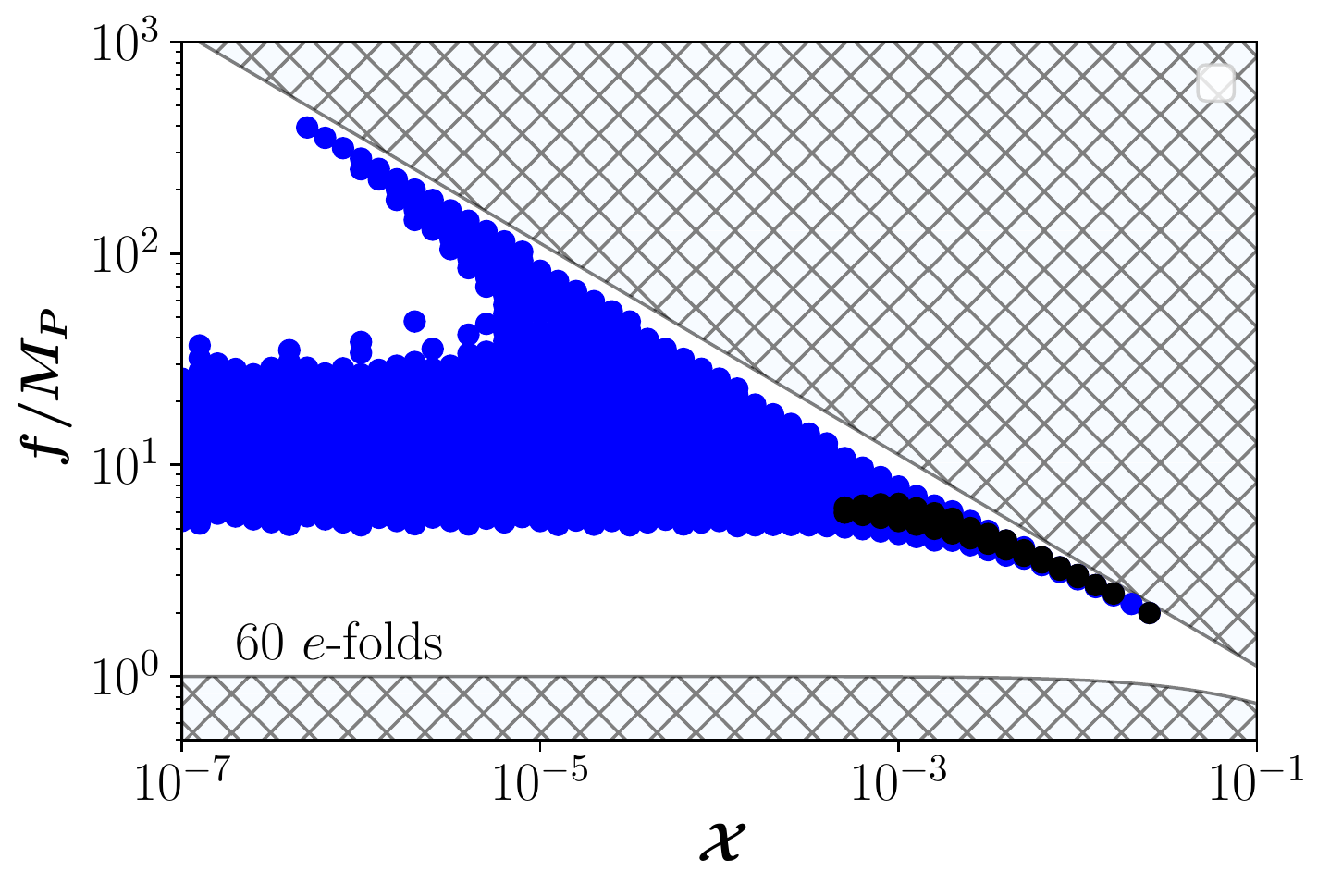}
	\caption{The blue and black points correspond to the parameter space compatible with the Planck observation at $95\%$ and $68\%$~CL, respectively. Left (right) panel shows the case of 50 (60) $e$-folds.
	}
	\label{fig:f-xi}
\end{figure}
Fig.~\ref{fig:f-xi} depicts the values of $f$ and $\xi$ compatible with Planck limits at $68\%$ (black dots) and $95\%$ CL (blue dots), for positive $\nm$.
Left (right) panel shows the case of 50 (60) $e$-folds.
The hashed regions correspond to the parameter space where $N_e$ and $n_s$ tend to be negative at small field values, respectively, and are therefore disregarded.
We note that in order to agree with Planck measurements, the best fit favors small but finite non-minimal couplings of the order $\mathcal{O}(10^{-3})$ to $\mathcal{O}(10^{-2})$, and $f$ at the Planck scale.

\subsection{Naturally Steep Potential with Non-minimal Coupling to Gravity} 
In this case where we have simultaneously a non-minimal coupling to gravity and interactions with gauge fields, the slow-roll parameter $\epsilon$ can be approximate as
\ba
\bar{\epsilon} = -\frac{ \dot{\bar{H}} }{ \bar{H}^2 } &\approx& \frac{    3 \Omega \,K\, \dot{\phi}^2  }{   {2V}     } + \frac{      \langle \vec{E}^2 + \vec{B}^2 \rangle   }{   {V}     } .
\ea
Taking into account that 
\be
\bar{H}^2 = \frac{H^2}{\Omega} +  \frac{\dot{\Omega} H}{\Omega^2} + \frac{1}{4} \frac{\dot{\Omega}^2}{\Omega^3} \approx  \frac{H^2}{\Omega} ,
\ee
where we neglected the $\dot{\Omega}$ terms since they are proportional to $\dot{\phi}$, we get
\be 
\bar{\epsilon} =  3 \Omega\, \frac{K }{2} \,\frac{  4 \xi^2\, f^2\, H^2  }{ \alpha^2\, {V}  } + \frac{      \langle \vec{E}^2 + \vec{B}^2 \rangle   }{   {V}     } =    \frac{  2  K \,\xi^2 f^2  }{ \alpha^2 M_{P}^2  } + \frac{8}{7}\,\frac{\xi}{\alpha}\,\frac{f \,K^{1/2}\,\bar{V}_{\bar{\phi}}}{ \bar{V}}, 
\ee
where we have used Eq.~\eqref{Asol}. 
We also consider the slow-roll parameter $\bar{\eta}$
\be
\bar{\eta} = \bar{\epsilon} - \frac{1}{2  \bar{\epsilon} } \frac{d \bar{\epsilon}}{d \bar{N}} = \bar{\epsilon} - \frac{1}{2  \bar{H}\, \bar{\epsilon} } \frac{d \bar{\epsilon}}{d \bar{t}} =-\frac{ \frac{d^2 \bar{\phi}}{{d} \bar{t}^2 } }{ \frac{d \bar{\phi} }{d \bar{t}}\,  \bar{H} }\,. 
\ee

\begin{figure}[t!]
	\centering
	\includegraphics[width=0.47\textwidth]{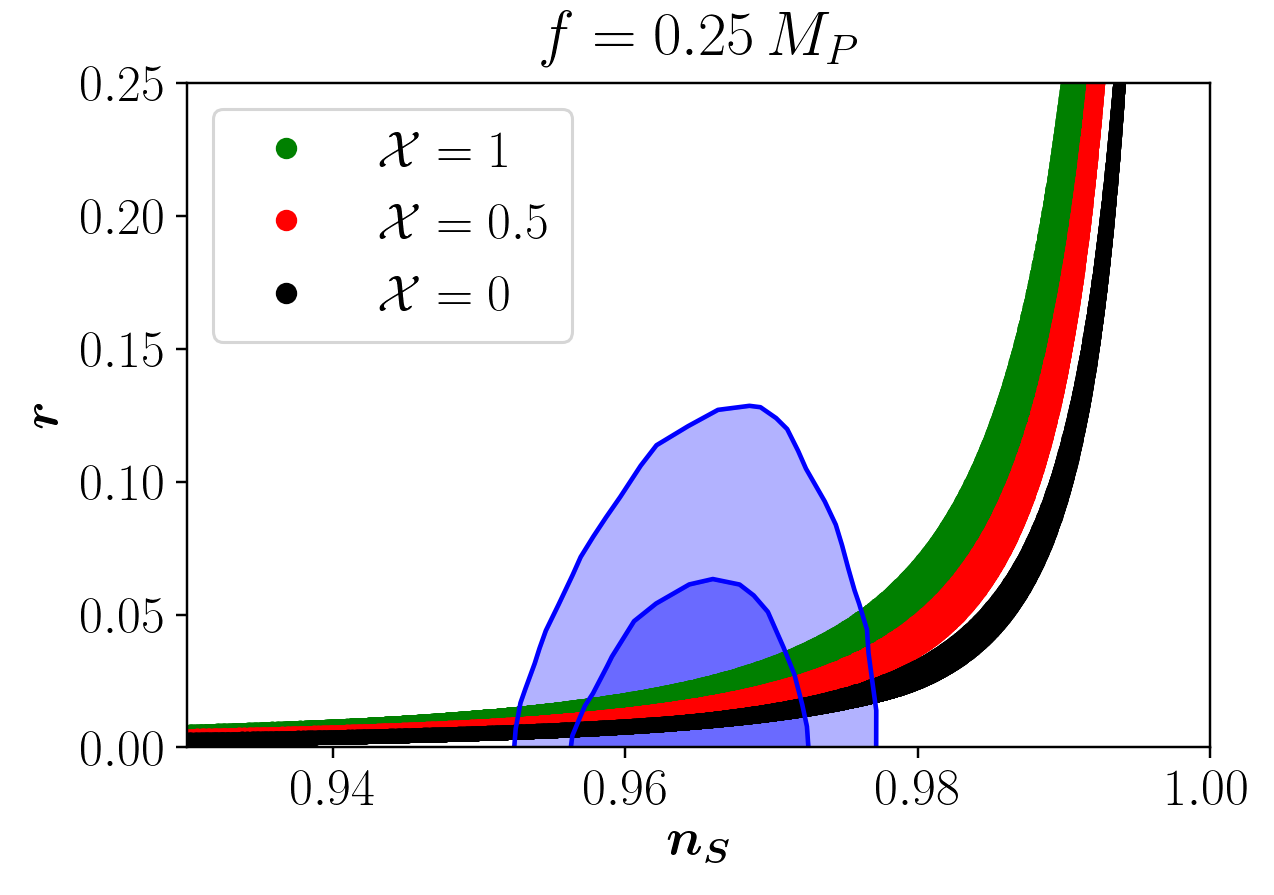}
	\includegraphics[width=0.47\textwidth]{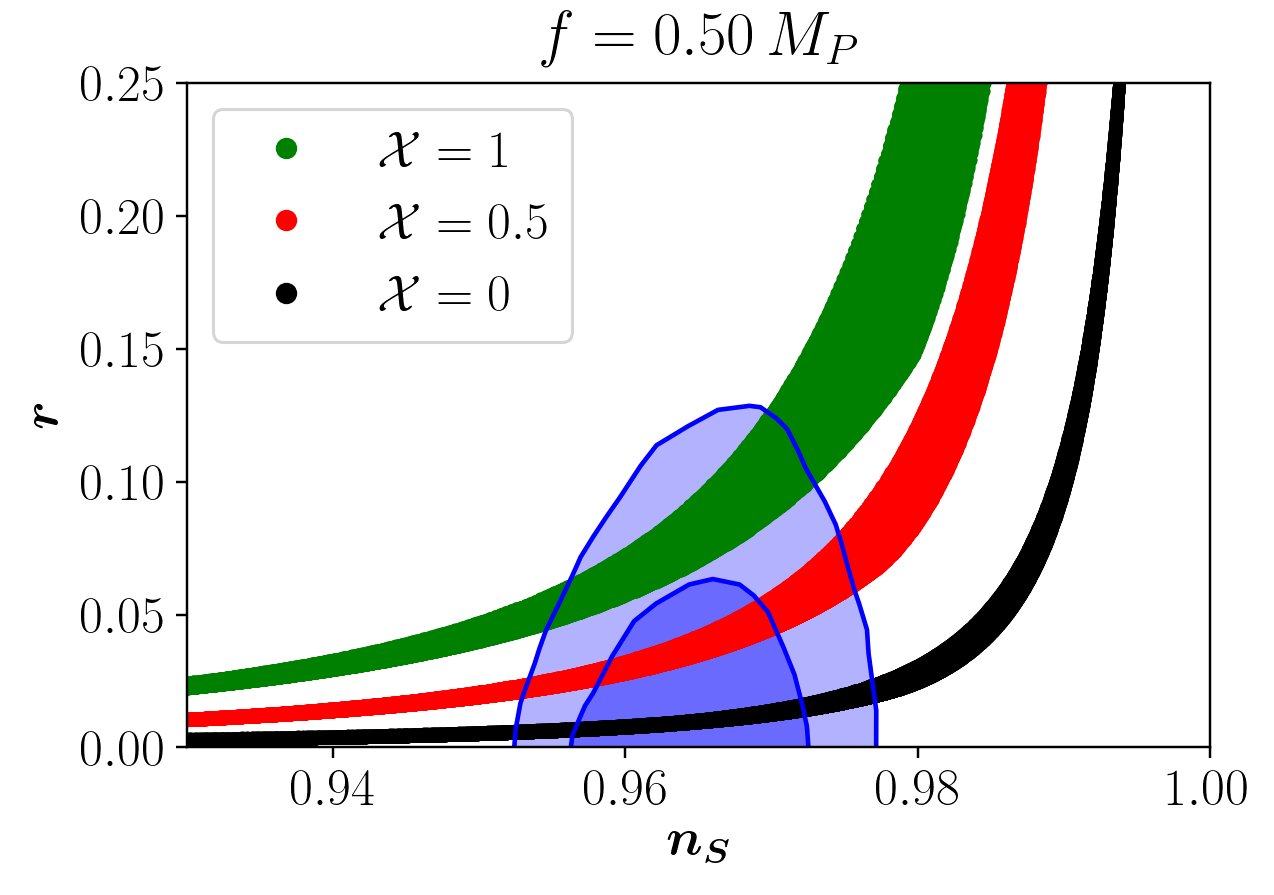}
	\caption{Tensor-to-scalar ratio $r$ versus the scalar spectral index $n_s$ for 60 $e$-folds, in the case where the inflaton is coupled with $\mathcal{N}$ gauge fields and non-minimally coupled to gravity.
	}
	\label{fig:r-ns_photon_nm}
\end{figure}
Fig.~\ref{fig:r-ns_photon_nm} is the equivalent of Fig.~\ref{fig:r-ns_photon} but now with the inclusion of non-minimal couplings.
It again shows the $68\%$ (light blue) and $95\%$ (dark blue) CL regions for the tensor-to-scalar ratio $r$ versus the scalar spectral index $n_s$ from Planck.
The predictions in the case of natural inflation with the inflaton non-minimally coupled to gravity and also coupled with $\mathcal{N}$ gauge fields are also shown, for $f=0.25\,M_P$ (left panel) and $f=0.5\,M_P$ (right panel), and $\nm=1$ (green), 0.5 (red) and 0 (black).
These regions were constructed by fixing $N_e=60$ but varying the parameters in the ranges: $\xi=[2.5,\,10]$ and $\alpha=[80,\,400]$.
The number $\cal N$ of gauge fields is fixed by the COBE normalization, the scale $\Lambda$ of the potential by the equation of motion of the inflaton.
In the case where $\nm=0$ (i.e. inflaton minimally coupled to gravity), the results are independent from $f$, and coincide with the ones of Fig.~\ref{fig:r-ns_photon}.
Higher values for $\nm$ tend to increase $r$, even above the preferred region by Planck.
That is particularly true for large values of $f$.
The values for $\xi$, $\alpha$, $\mathcal{N}$ and $\Lambda$ compatible with Planck limits are not presented here because they are basically the same as in the minimal case, i.e. Fig.~\ref{fig:xi_photon}.

\section{Conclusions and Final Remarks}
\label{Conclusions}
We studied a scenario in which a pseudoscalar field $\phi$ is coupled with an ensemble of gauge fields through the axial term $\phi \tilde{F}F$ in the presence of a non-minimal coupling to gravity. Due to the axial coupling, there is a significant production of chiral gravitational waves which would leave a characteristic signature in the CMB. However, there are strong constraints over this mechanism coming mainly from non-Gaussianity and perturbativity~\cite{Ferreira:2014zia,Ferreira:2015omg,Peloso:2016gqs}.\footnote{Nevertheless, as we mentioned in the introduction, here perturbativity of the background evolution is broken by construction. The scalar and tensor perturbations remain perturbative.} In this paper, we added a non-minimal coupling between the pseudoscalar inflaton and gravity, and studied the spectrum of the scalar and tensor perturbations. We tracked the effects of this non-minimal coupling on the spectral index of scalar perturbations $n_{s}$ and on the tensor-to-scalar ratio $r$, in order to find regions of the parameter space compatible with current observational bounds. The main results of this study are Eqs.~\eqref{psscalenm},~\eqref{nsnmc},~\eqref{rVV1nmc} and the scan of parameters represented in Fig.~\ref{fig:r-ns_photon_nm}. We found that, as a result of the gauge fields and gravity interactions with the pseudoscalar inflaton for a steep cosine potential, $n_s$ and $r$ are typically mildly modified. However, there are also regions of the parameters space that present a suppression of the vacuum gravitational waves accompanied with an enhancement of the gravity waves produced by the axial coupling interaction. This enhancement of the sourced gravitational waves happens because the dynamics of the gauge fields is affected by the non-minimal coupling through the function $K$ in Eq.~\eqref{Kfunction} and the parameter $\xi$ in Eq.~\eqref{apm}. The $K$ function allows for significant modifications  in the sourced gravitational wave term, while $\xi$ is almost unaltered. 

An interesting possibility offered by the interaction with gauge fields and a non-minimal coupling is that the amplification of gravitational waves could alleviate some tension with non-Gaussianity bounds.
In fact, it was pointed out that non-Gaussianity strongly constrains axionic models with slow-roll potential~\cite{Ferreira:2014zia}, requiring $\xi\lesssim 2.5$. For such small values, the gravitational waves generated by the the axial coupling are comparable to the vacuum ones, and hence practically unobservable. 
However the situation here is different because the non-minimal coupling modifies the evolution of the parameter $\xi$ and the slow-roll parameter $\bar{\epsilon}$ for the canonical variables. We will follow the steps of the reasoning of Ref.~\cite{Ferreira:2014zia} to track the modifications due to the non-minimal coupling.   
In the same way that the scalar spectrum is unaltered in the Einstein frame, the bispectrum and hence the non-Gaussianity parameter $\bar{f}_{\rm NL}$ do not change either, so 
\be
\bar{f}_{\rm NL}^{\rm equil} \approx f_{\rm NL}^{\rm equil} \approx 4.4\times 10^{10} {\cal \bar{P}_{\zeta}}^3 \frac{e^{6\pi \xi}}{\xi^9}.
\ee
With this, we get the bound  
\be
 {\cal \bar{P}_{\zeta}}^2 \frac{e^{4\pi \xi}}{\xi^6} < \left(\frac{ f_{\rm NL}^{\rm obs}}{ 4.4\times 10^{10}} \right)^{2/3}.
\ee
Inserting this expression in Eq.~\eqref{barr} we obtain
\be
\bar{r} \approx \frac{{\cal A}^{+}   \bar{H}^4}{\pi^2 M_P^4} \frac{  \frac{e^{4\pi \xi}}{\xi^6} }{\bar{{\cal P}}_{\zeta}} < 35  \frac{ \bar{H}^4}{ M_P^4}   \frac{e^{4\pi \xi}}{\xi^6} < 1.7\times 10^{-2}\left( f_{\rm NL}^{\rm obs} \right)^{2/3} \bar{\epsilon}^2.
\ee
In the last expression, $\bar{\epsilon} $ is the slow-roll parameter for the canonical field $\bar{\phi}$. It can be seen that this parameter is related with the $\epsilon$ parameter for the non canonical field $\phi$ through 
\be
\epsilon = \frac{\dot{\phi}^2}{2 H^2} \approx \frac{1}{K}\frac{  \left( \frac{d\bar{\phi}}{d\bar{t}} \right)^2 }{2 \bar{H}^2} =  \frac{1}{K} \bar{\epsilon},
\ee 
where we used Eq.~\eqref{attauOm}. With that, we rewrite the bound as
\be
\bar{r}  < 1.7\times 10^{-2}\left( f_{\rm NL}^{\rm obs} \right)^{2/3} K^2 {\epsilon}^2.
\ee
Here we notice that the slow-roll parameter $\epsilon$ for the non-canonical field is not modified significantly due to the non-minimal coupling function. However, the function $K$ can act as an amplification factor for certain values of the non-minimal coupling parameter $\nm$.
This mechanism offers then a possibility to leave some measurable signatures of sourced gravitational waves.

\section*{Acknowledgments}
We acknowledge Lorenzo Sorbo for his valuable correspondence, Marta Losada for our useful discussions and Juan Carlos Bueno S\'anchez for the early discussions on the topics considered here.
JPBA acknowledges partial funding from COLCIENCIAS grants numbers 123365843539 RC FP44842-081-2014 and 110671250405 RC FP44842-103-2016.
NB is partially supported by the Spanish MINECO under Grant FPA2017-84543-P.
This project has received funding from the European Union's Horizon 2020 research and innovation programme under the Marie Skłodowska-Curie grant agreements 674896 and 690575; and from Universidad Antonio Nari\~no grants 2017239 and 2018204.

\appendix

\section{Details of the Calculation of the Perturbations Spectrum}\label{AA}
\subsection{The Green Function}
Here we revisit the details of the approximate solution for the scalar perturbations. 
The RHS of Eq.~\eqref{phipert} can be decomposed in two parts, one due to the intrinsic inhomogeneities in $\vec{E}\cdot \vec{B}$ even with $\phi =0$, and a second component due to the dependence on $\dot{\phi}$: 
\be
\delta\left[\vec{E}\cdot \vec{B} \right] \approx \left[\vec{E}\cdot \vec{B} - \langle \vec{E}\cdot \vec{B} \rangle  \right]_{\delta\phi =0} + \frac{\partial  \langle \vec{E}\cdot \vec{B} \rangle }{\partial \dot{\phi} } \delta  \dot{\phi} \equiv \delta_{\vec{E}\cdot \vec{B} }+ \frac{\partial  \langle \vec{E}\cdot \vec{B} \rangle }{\partial \dot{\phi} } \delta  \dot{\phi}\,.
\ee
Using the definition of $\xi$ and the approximations in Eq.~\eqref{EBxi}, the $\dot{\phi}$ dependent term can be written as 
\be
 \frac{\partial  \langle \vec{E}\cdot \vec{B} \rangle }{\partial \dot{\phi} } \delta  \dot{\phi} \approx  \frac{\partial  \langle \vec{E}\cdot \vec{B} \rangle }{\partial \xi }  \frac{\partial  \xi }{\partial \dot{\phi} }  \delta \dot{\phi} \approx  \frac{\partial  \langle \vec{E}\cdot \vec{B} \rangle }{\partial \xi }  \frac{\alpha }{2 f H }  \delta \dot{\phi} \approx  2\pi \langle \vec{E}\cdot \vec{B} \rangle \frac{\alpha }{2 f  {a} H }  \delta {\phi}',
\ee
where we have used $dt = a\, d\tau$ in the last part. Dots and primes correspond to $t$ and $\tau$ derivatives, respectively. If we use the homogeneous Eq.~\eqref{phi0}, and assuming that the dynamics is governed by the source term, we can neglect the time derivatives of the pseudoscalar field and hence 
\be \label{vphieb}
V_{\phi} \approx \frac{ {{\cal N}}\alpha}{f} \langle \vec{E}\cdot \vec{B} \rangle.
\ee
Putting everything together, we write the perturbations equation as
\ba\label{pertback}
 \delta\phi'' + 2aH\delta\phi' +\left[ - \nabla^2 +  a^2 {V_{\phi\phi}(\phi_0)} \right]\delta\phi  \approx  a^2 \frac{  {{\cal N}}\alpha}{f} \left(\delta_{\vec{E}\cdot \vec{B} }  +  \frac{ \pi V_{\phi}(\phi_0) }{  {{\cal N}}   {a} H }  \delta {\phi}'  \right)\,,\\
 \delta\phi'' + 2aH \left( 1  {-}  \frac{  \alpha \pi V_{\phi}(\phi_0)}{2fH^2} \right) \delta {\phi}' + \left[ - \nabla^2 +  a^2 {V_{\phi\phi}(\phi_0)} \right] \delta\phi  \approx  a^2 \frac{  {{\cal N}}\alpha}{f} \delta_{\vec{E}\cdot \vec{B} }\,.
\ea
Using $a \approx  {-}1/(H\tau)$ during inflation (de Sitter metric), we get
\be\label{A6}
 \delta\phi''   {-} \frac{2}{\tau} \left( 1  {-}  \frac{  \alpha \pi V_{\phi}(\phi_0)}{2fH^2} \right) \delta {\phi}' +\left[ - \nabla^2 +   \frac{1}{\tau^2} \frac{V_{\phi\phi}(\phi_0)}{H^2} \right] \delta\phi  \approx  a^2 \frac{  {{\cal N}}\alpha}{f} \delta_{\vec{E}\cdot \vec{B} }.
\ee
Assuming that the inhomogeneities are small, and considering also that $V_{\phi} \approx V/f$ and $V\approx 3M_P^2\, H^2$ and assuming $\alpha \gg 1$ and $f\lesssim M_P$, the factor accompanying $\delta \phi'$ can be approximated as
\be
  {-} \frac{2}{\tau}\left( 1  {-}  \frac{  \alpha \pi V_{\phi}(\phi_0)}{2fH^2} \right)  \approx    {+} \frac{1}{\tau}  \frac{  \alpha \pi V_{\phi}(\phi_0)}{fH^2}.
\ee
The last approximation is valid in this case where the backreaction term, the one coming from the variation  $\delta[\vec{E}\cdot \vec{B}]$ in the RHS of Eq.~\eqref{pertback} is the dominant.
In Eq.~\eqref{A6}, the gradient can also be neglected.
In fact, the bulk of the analysis relies on the solution for the vector potential $A^\mu$, Eq.~\eqref{Asol}, which is valid in the regime $|k \tau| \ll 2\,\xi$, with $\xi > 1$.
Considering the third term in Eq.~\eqref{A6}, one can compare the relevance of the gradient term with the friction term.
It can be approximated by
\begin{equation}
k^2 + \frac{1}{\tau^2}\frac{V''}{H^2} \approx k^2 + \frac{1}{\tau^2}\frac{V}{f^2 H^2 } \approx k^2 + \frac{3\,M_P^2}{\tau^2f^2}\,.
\end{equation}
We can therefore neglect the gradient term whenever $|k\tau|\lesssim M_P/f$.
Taking into account that $|k \tau| \ll 2\,\xi$, we deduce that this condition is always satisfied as long as $f\lesssim M_P/\xi$:
subplanckian values for the coupling $f$, making the potential steep, are required in this mechanism and ensure that the electromagnetic friction term dominates over the gradient of the scalar potential.

Therefore, Eq.~\eqref{A6} becomes 
\be \label{pertphiappA}
 \delta\phi''   {+} \frac{1}{\tau}   \frac{  \alpha \pi V_{\phi}(\phi_0)}{fH^2}\delta {\phi}' +    \frac{1}{\tau^2} \frac{V_{\phi\phi}(\phi_0)}{H^2} \delta\phi  \approx  a^2 \frac{  {{\cal N}}\alpha}{f} \delta_{\vec{E}\cdot \vec{B} }.
\ee
We can solve the Green's function for this equation:
\be
 \frac{\partial^2G (\tau ,\tau')}{\partial \tau^2}   {+} \frac{1}{\tau}   \frac{  \alpha \pi V_{\phi}(\phi_0)}{fH^2}  \frac{\partial G (\tau, \tau')}{\partial \tau} +   \frac{1}{\tau^2} \frac{V_{\phi\phi}(\phi_0)}{H^2} G (\tau, \tau')  =  \delta(\tau - \tau') 
\ee
with the boundary conditions $G (\tau',\tau' ) = 0$, $\partial G (\tau',\tau' )/\partial\tau =1$, obtaining
\be\label{green}
G (\tau, \tau') = \frac{\tau'}{\Delta}\left[ \left( \frac{\tau}{\tau'} \right)^{\nu_{+}} -\left( \frac{\tau}{\tau'} \right)^{\nu_{-}}\right]\Theta(\tau - \tau')\,,
\ee
where
\be 
\nu_{\pm} \equiv \frac{1}{2}\left(1-\frac{\pi\,\alpha\, V_{\phi}(\phi_0)}{f\,H^2} \right) \pm \frac{1}{2}\Delta, \quad \mbox{and} \quad \Delta \equiv \sqrt{\left(1-\frac{\pi\,\alpha\, V_{\phi}(\phi_0)}{f\,H^2} \right)^2 - \frac{4 V_{\phi\phi}(\phi_0)}{H^2}}\,.
\ee
As said before, $\frac{\pi\alpha V_{\phi}(\phi_0)}{fH^2} \gg 1$, then we can approximate the previous expressions as 
\be
\Delta \approx \frac{\pi\alpha V_{\phi}(\phi_0)}{fH^2}\sqrt{ 1 - \frac{4 f^2 H^2 V_{\phi\phi}(\phi_0)}{ \pi^2 \alpha^2 V_{\phi}(\phi_0)^2 }},\quad \nu_{\pm} \approx -\frac{1}{2}\frac{\pi\alpha V_{\phi}(\phi_0)}{fH^2} \left( 1 \mp \sqrt{ 1 - \frac{4 f^2 H^2 V_{\phi\phi}(\phi_0)}{ \pi^2 \alpha^2 V_{\phi}(\phi_0)^2 }} \right).
\ee
Taking into account that $[f/(\alpha M_P)]^2 \ll 1$, previous equations can be rewritten as
\be\label{rootsapp} 
\Delta \approx \frac{\pi\alpha V_{\phi}(\phi_0)}{fH^2},\quad  \nu_{+} \approx  {-}\frac{ f V_{\phi\phi}(\phi_0)}{ \pi\alpha V_{\phi}(\phi_0) }\ll1, \quad \nu_{-} \approx -\frac{\pi\alpha V_{\phi}(\phi_0)}{fH^2}\approx -\Delta\gg1\,.
\ee
At late times or for scales at which $|k\tau|\ll 1$, the Green's function becomes
\be\label{Greenapprox}
G (\tau, \tau') \approx \frac{\tau'}{\Delta} \left( \frac{\tau}{\tau'} \right)^{\nu_{+}} , \quad \tau >\tau'.
\ee

\subsection{Power Spectrum}
We need to calculate the spectrum of the scalar perturbations: 
\ba\nonumber
\langle \delta \phi(\vec{p}) \delta \phi(\vec{p}\,') \rangle &=& \delta(\vec{p} + \vec{p}') \left(\frac{ {{\cal N}} \alpha }{f}\right)^2\int d\tau_1 d\tau_2 a_1^2 a_2^2 G(\tau, \tau_1)G(\tau, \tau_2) \\
&& \qquad \qquad \times \int d^3 x e^{i\vec{p}\cdot\vec{x}}\langle \delta_{\vec{E}\cdot \vec{B} }(\tau_1, 0) \delta_{\vec{E}\cdot \vec{B} }(\tau_2, \vec{x})\rangle.
\ea
Using Eq.~\eqref{Greenapprox} we can approximate the previous expression to
\be\label{psapprox}
\langle \delta \phi(\vec{p}) \delta \phi(\vec{p}\,') \rangle \approx \delta(\vec{p} + \vec{p}\,') \frac{  {{\cal N}^2} \alpha^2  {\tau^{2\nu_{+}}} }{H^4 \Delta^2 f^2}\int  \frac{ d\tau_1 d\tau_2 }{ (\tau_1 \tau_2)^{1+\nu_{+}}  } \int d^3 x e^{i\vec{k}\cdot\vec{x}}\langle \delta_{\vec{E}\cdot \vec{B} }(\tau_1, 0) \delta_{\vec{E}\cdot \vec{B} }(\tau_2, \vec{x})\rangle.
\ee
The factor $  {\tau^{2\nu_{+}}}$ carries the scale dependence in the power spectrum. Following Refs.~\cite{Anber:2009ua, Anber:2012du} we  evaluate the spacial integral using the approximation (\ref{Asol}), and neglecting the $A_{-}$ polarization one finds
\ba
\int d^3 x e^{i\vec{k}\cdot\vec{x}}\langle \delta_{\vec{E}\cdot \vec{B} }(\tau_1, 0) \delta_{\vec{E}\cdot \vec{B} }(\tau_2, \vec{x})\rangle &=& \frac{e^{4\pi \xi} }{4a_1^4 a_2^4} \int \frac{d^3 k}{(2\pi)^3} | \epsilon_{+}(-\vec{k}) \cdot  \epsilon_{+}(\vec{p}+\vec{k}) |^2  \\ \nonumber
&\times& \left[ |\vec{k}| |\vec{k} + \vec{p} |  + |\vec{k}|^{3/2} |\vec{k} + \vec{p} |^{1/2} \right] e^{-4\sqrt{ 2\xi/\tilde{a}H } \left( \sqrt{ |\vec{k}| } + \sqrt{ |\vec{k}+\vec{p} | }\right) } 
\ea
where $2/\sqrt{\tilde{a}} \equiv 1/\sqrt{a_1} + 1/\sqrt{a_2}$. We can choose the vector $\vec{p}$ to be along the $\hat{z}$ axis, then $\vec{p} = p\, \hat{z}$, and we change variables to  $\vec{q}\equiv \vec{k}/p$. With this, the integral is expressed as
\ba\label{intdeltadelta}
\int d^3 x e^{i\vec{k}\cdot\vec{x}}\langle \delta_{\vec{E}\cdot \vec{B} }(\tau_1, 0) \delta_{\vec{E}\cdot \vec{B} }(\tau_2, \vec{x})\rangle &=& \frac{e^{4\pi \xi}  {p^5} }{4a_1^4 a_2^4} \int \frac{d^3 q}{(2\pi)^3} | \epsilon_{+}(-\vec{q}) \cdot  \epsilon_{+}(\hat{z}+\vec{q}) |^2  \\ \nonumber
&\times& q  |\hat{z} + \vec{q} | \left[ 1  + \frac{q^{1/2}}{ |\hat{z} + \vec{q} |^{1/2}} \right] e^{-\sqrt{ 2^5\xi  {p}/\tilde{a}H } \left( \sqrt{ |\vec{q}| } + \sqrt{ |\vec{q}+\hat{z} | }\right) } .
\ea
The time dependence of the exponential factor is separated as
\be
e^{-\sqrt{ 2^5\xi  {p}/\tilde{a}H } \left( \sqrt{ |\vec{q}| } + \sqrt{ |\vec{q}+\hat{z} | }\right) } = e^{-\frac{1}{2}\left( \sqrt{ -2^5\xi  {p}\tau_1 } + \sqrt{ -2^5\xi  {p} \tau_2} \right)  \left( \sqrt{ |\vec{q}| } + \sqrt{ |\vec{q}+\hat{z} | }\right) } 
\ee
which, introducing the change of variables $x_{1,\,2} =(-2^5\xi \, {p} \,\tau_{1,\,2})^{1/4}$ becomes
\be
e^{-\frac{1}{2}\left( \sqrt{ -2^5\xi  {p}\tau_1 } + \sqrt{ -2^5\xi  {p} \tau_2} \right)  \left( \sqrt{ |\vec{q}| } + \sqrt{ |\vec{q}+\hat{z} | }\right) } = e^{-\frac{1}{2}\left( x_1^2 + x_2^2 \right)  \left( \sqrt{ |\vec{q}| } + \sqrt{ |\vec{q}+\hat{z} | }\right) } ,
\ee
and the integral (\ref{intdeltadelta}) is expressed as
\ba
\int d^3 x e^{i\vec{k}\cdot\vec{x}}\langle \delta_{\vec{E}\cdot \vec{B} }(\tau_1, 0) \delta_{\vec{E}\cdot \vec{B} }(\tau_2, \vec{x})\rangle &=& \frac{e^{4\pi \xi}  {p^5} H^8 x_1^{16} x_2^{16} }{ (2^5 \xi  {p})^8  2^5  \pi^3} {\cal M } (x_1, x_2)\,,
\ea
where
\ba
{\cal M } (x_1, x_2) &\equiv &  \int {d^3 q} {\cal J}(\vec{q}) e^{-\frac{1}{2}\left( x_1^2 + x_2^2 \right)  \left( \sqrt{ |\vec{q}| } + \sqrt{ |\vec{q}+\hat{z} | }\right) } \\ \nonumber
  & \equiv & \int {d^3 q} | \epsilon_{+}(-\vec{q}) \cdot  \epsilon_{+}(\hat{z}+\vec{q}) |^2   q  |\hat{z} + \vec{q} | \left[ 1  + \frac{q^{1/2}}{ |\hat{z} + \vec{q} |^{1/2}} \right] e^{-\frac{1}{2}\left( x_1^2 + x_2^2 \right)  \left( \sqrt{ |\vec{q}| } + \sqrt{ |\vec{q}+\hat{z} | }\right) } .
\ea
Inserting this into Eq.~\eqref{psapprox} and doing the time variables change, we get
\be\label{psapproxom}
\langle \delta \phi(\vec{p}) \delta \phi(\vec{p}\,') \rangle = \delta(\vec{p} + \vec{p}\,') \frac{  {{\cal N}^2} \alpha^2  H^4 }{ {p^3}   \Delta^2 f^2 \xi^8 }  \frac{e^{4\pi \xi} }{2^{15}}  { (-2^5 \xi p \tau)^{2\nu_{+}}} \int  dx_1 dx_2  (x_1 x_2)^{15 -4\nu_{+}  }   \frac{ 16 {\cal M } (x_1, x_2)  }{    2^{30}  \pi^3}. 
\ee
The time integrals in the variables $x_i$ can be performed as integrals of the form $x^{\beta} e^{-\alpha x^2}$ 
 \ba\label{timeint}\nonumber
&&\int  dx_1 dx_2  (x_1 x_2)^{15 -4\nu_{+}  }  \frac{  16 {\cal M } (x_1, x_2)  }{    2^{30}  \pi^3}\nn\\ 
&&\hspace{2cm}=   \int {d^3 q} {\cal J}(\vec{q}) \frac{16}{ 2^{30}  \pi^3}\int  dx_1 dx_2 (x_1 x_2)^{15 -4\nu_{+}  }   e^{- \frac{1}{2}\left( \sqrt{ |\vec{q}| } + \sqrt{ |\vec{q}+\hat{z} | }\right) \left( x_1^2 + x_2^2\right)} \nn\\
&&\hspace{2cm}=  \frac{ \Gamma[8 -2 \nu_{+} ]^2}{2^{ 12+4 \nu_{+} } \pi^3} \int {d^3 q} \frac{ {\cal J}(\vec{q}) }{ \left( \sqrt{ |\vec{q}| } + \sqrt{ |\vec{q}+\hat{z} | }\right)^{16 -4 \nu_{+}} }   . 
\ea
Using 
\be
 | \epsilon_{+}(-\vec{q}) \cdot  \epsilon_{+}(\hat{z}+\vec{q}) |^2 = \frac{1}{4}\left( \frac{q +\cos{\theta} -  |\vec{q}+\hat{z} | }{  |\vec{q}+\hat{z} | } \right)^2,
\ee
where $\theta$ is the angle between $\hat{z}$ and $\vec{q}$, the momentum integral becomes 
 \be
\int {d^3 q} \frac{ {\cal J}(\vec{q}) }{ \left( \sqrt{ |\vec{q}| } + \sqrt{ |\vec{q}+\hat{z} | }\right)^{16 -4 \nu_{+}} }= \int {d^3 q} \frac{ \left( {q +\cos{\theta} -  |\vec{q}+\hat{z} | } \right)^2   q  }{ 4  |\vec{q}+\hat{z} |\left( \sqrt{ |\vec{q}| } + \sqrt{ |\vec{q}+\hat{z} | }\right)^{16 -4 \nu_{+}} } \left[ 1  + \frac{q^{1/2}}{ |\hat{z} + \vec{q} |^{1/2}} \right]   . 
\ee
Then the momentum integral can be evaluated numerically. Neglecting $\nu_{+}\ll 1$ we obtain
 \ba
 \int {d^3 q} \frac{ {\cal J}(\vec{q}) }{ \left( \sqrt{ |\vec{q}| } + \sqrt{ |\vec{q}+\hat{z} | }\right)^{16 -4 \nu_{+}} }\approx 3.49\times 10^{-4}   . 
\ea
And the time integral is evaluated as
\ba
{\cal T}(\nu_{+}=0) = \int  dx_1 dx_2  (x_1 x_2)^{15 -4\nu_{+}  }  \frac{  16 {\cal M } (x_1, x_2)  }{    2^{30}  \pi^3} \approx  3.49\times 10^{-4}  \frac{ \Gamma[8]^2}{2^{ 12 } \pi^3} \approx 7\times 10^{-2}. \quad
\ea
Notice that, if we use  $\nu_{+} \approx -0.0175$, to be in agreement with Planck results for the spectral index, the result is modified in a small amount 
\ba
{\cal T}(\nu_{+}\approx -0.0175 ) = \int  dx_1 dx_2  (x_1 x_2)^{15 -4\nu_{+}  }  \frac{  16 {\cal M } (x_1, x_2)  }{    2^{30}  \pi^3} \approx   8.2\times 10^{-2}. 
\ea
Using our previous results, the two point correlator of the perturbations is
\ba 
\langle \delta \phi(\vec{p}) \delta \phi(\vec{p}\,') \rangle & \approx & \frac{ {\cal T}(\nu_{+} ) }{2^{15}} \frac{\delta(\vec{p} + \vec{p}\,')}{p^3} \frac{  {{\cal N}^2} \alpha^2  H^4 }{   \Delta^2 f^2 \xi^8 }  e^{4\pi \xi}  { (-2^5 \xi p \tau)^{2\nu_{+}}} \nn\\
&\approx & {\cal F(\nu_{+})}  \frac{\delta(\vec{p} + \vec{p}\,')}{p^3} \frac{  {{\cal N}^2} \alpha^2  H^4 }{   \Delta^2 f^2 \xi^8 }  e^{4\pi \xi}  { (-2^5 \xi p \tau)^{2\nu_{+}}},
\ea
with ${\cal F}(\nu_{+}=0)\approx2.13 \times 10^{-6}$. With the previous results  we can calculate the power spectrum of the primordial curvature perturbation $\zeta = -H \delta \phi /\dot{\phi}_0$
\be
\delta(\vec{p} + \vec{p}\,'){\cal P}_{\zeta}(p) = \frac{p^3}{2 \pi^2} \frac{H^2}{\dot{\phi}_0^2} \langle \delta \phi(\vec{p})  \delta \phi(\vec{p}\,') \rangle \approx \delta(\vec{p} + \vec{p}\,')    \frac{{\cal F}}{2 \pi^2} \frac{H^2}{\dot{\phi}_0^2} \frac{  {{\cal N}^2} \alpha^2  H^4 }{   \Delta^2 f^2 \xi^8 }  e^{4\pi \xi}  { (-2^5 \xi p \tau)^{2\nu_{+}}}.
\ee
Using Eqs.~\eqref{apm}, \eqref{EBxi}, \eqref{phi0} and~\eqref{rootsapp}, the spectrum can be rewritten as
\be\label{psscale}
{\cal P}_{\zeta}(p) \approx    \frac{{\cal F} (\nu_{+})}{8 \pi^4 {\cal I}^2 \xi^2}    { (-2^5 \xi p \tau)^{2\nu_{+}}} \approx  \frac{4.7\times 10^{-2}}{  {\cal N}\xi^2}    { (-2^5 \xi p \tau)^{2\nu_{+}}} , 
\ee
where the factor  ${\cal N}$ comes from the fact that the gauge fields add incoherently in the spectrum.  From this expression we extract the spectral index
\be
n_s -1 \approx   { 2\nu_{+} } =  {-}\frac{2 f \,V_{\phi\phi}(\phi_0)}{ \pi\,\alpha\, V_{\phi}(\phi_0) } .
\ee

\section{Constant $\boldsymbol{\xi}$ regime}\label{constantxi}
\begin{figure}[t!]
	\centering
	\includegraphics[width=0.35\textwidth]{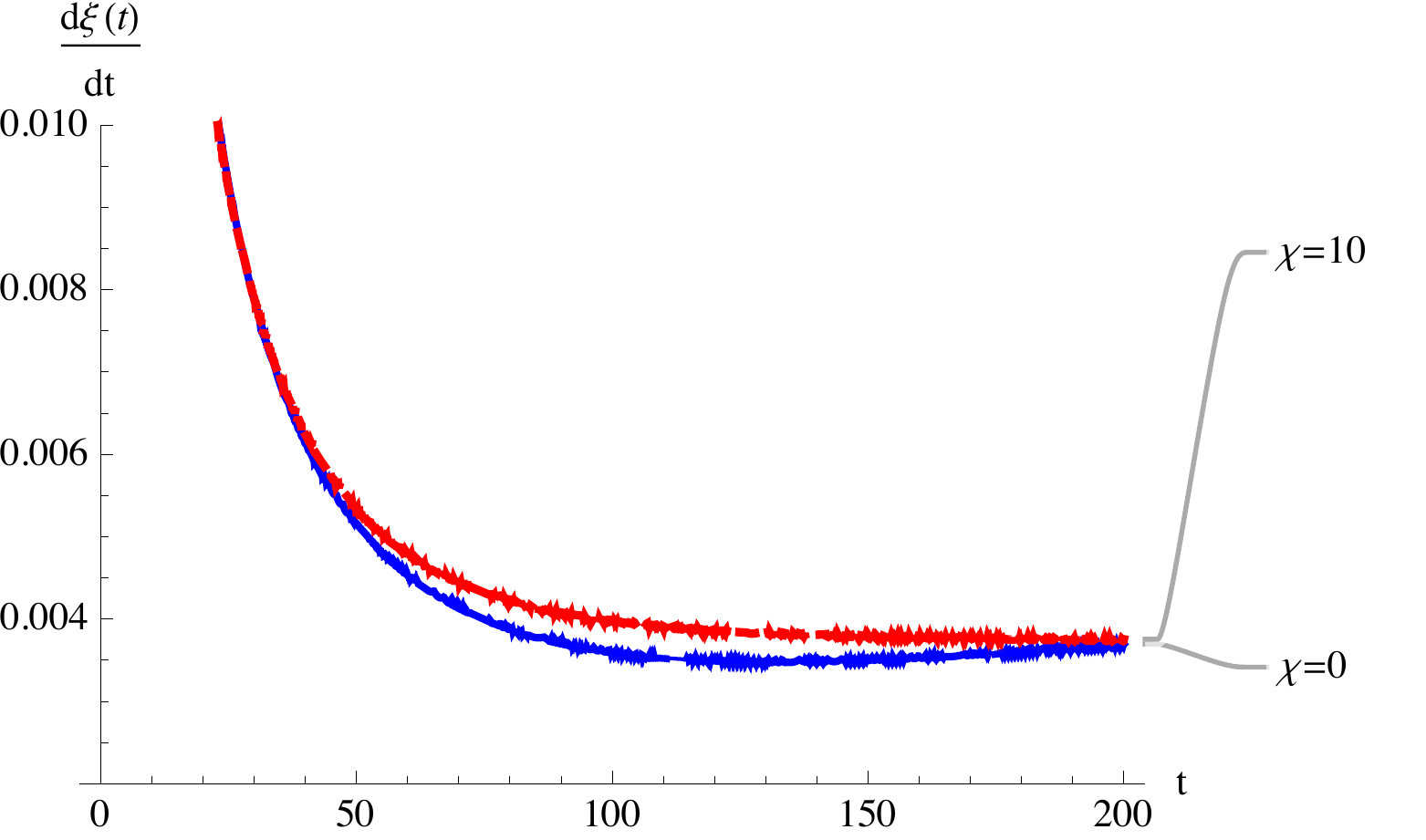}
	\includegraphics[width=0.35\textwidth]{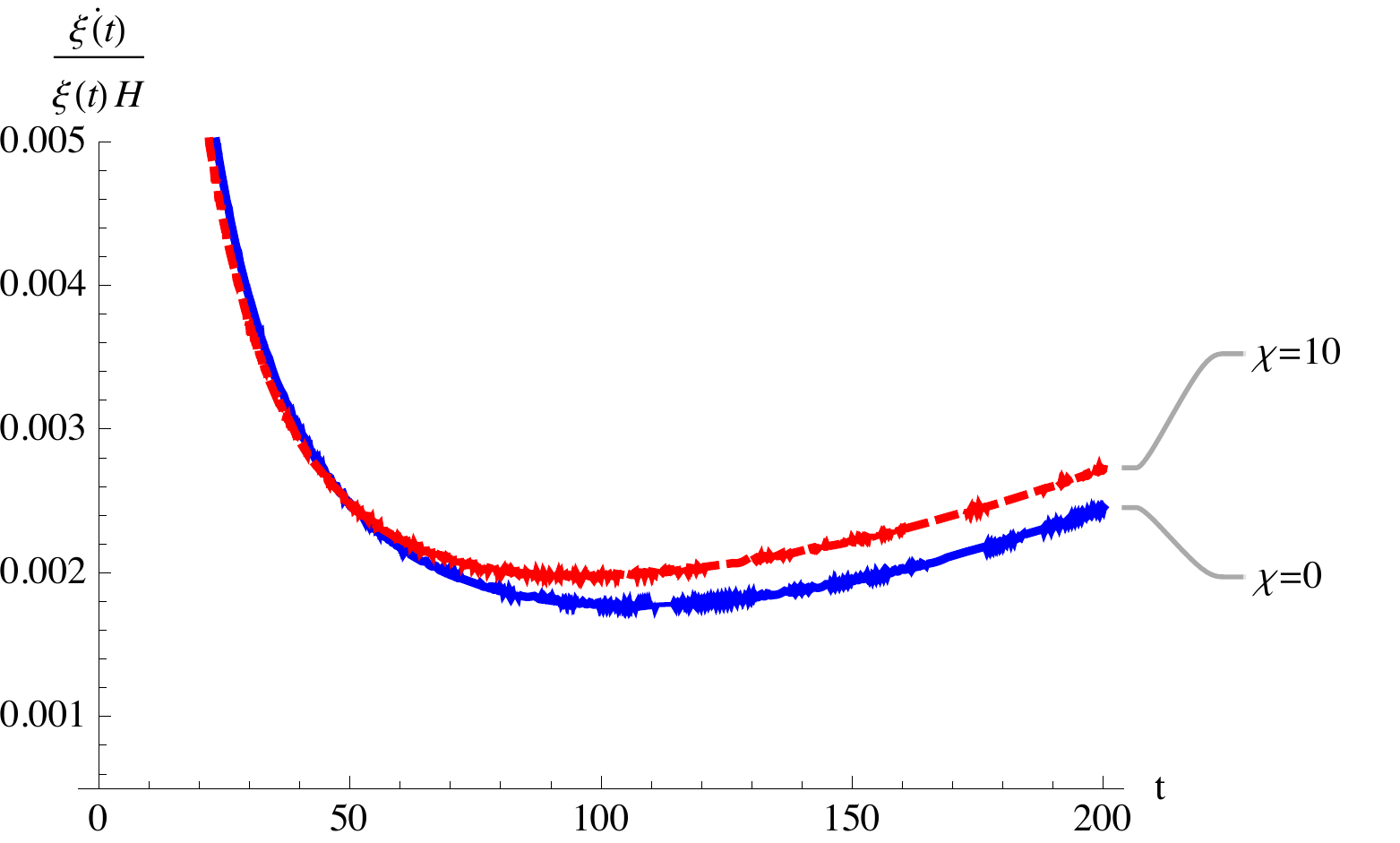}
\caption{In this figure we see the evolution of the following quantities: $\dot{\xi}$ at the  left panel, and $\dot{\xi}/(H\xi)$ at the right panel. The values used for those numerical evaluations are the same employed in Fig.~\ref{fig:xi_0and10} this is $\Lambda=4.5\times 10^{-3}M_{P}$, $\alpha=400$, ${\cal N}=10^5$, $f=0.1\,M_{P}$, and for minimal coupling (blue) and non-minimal coupling with $\nm = 10$ (red).  The time is measured in units of $M_{P}/\Lambda^2$.
	}
	\label{fig:constantxi_0and10}
\end{figure}

In this section we justify the assumption of constant $\xi$ used throughout the paper.
Fig.~\ref{fig:constantxi_0and10} shows the evolution of $\dot{\xi}$ and the ratio $\dot{\xi}/(H\xi)$, for the same benchmark point used in Fig.~\ref{fig:xi_0and10} ($\Lambda=4.5\times 10^{-3}M_{P}$, $\alpha=400$, ${\cal N}=10^5$, $f=0.1\,M_{P}$) and taking $\nm=0$ (blue) and $\nm=10$ (red).
Both quantities are small, which shows that the assumption of constant $\xi$ is well justified for the scales and the regime considered in the calculation of the perturbations. 

Additionally, the right panel shows that the ratio $\dot{\xi}/(H\xi)$ tends to stabilize around $2.5\times 10^{-3}$ for the time of horizon exit.
As the dependence of the solution for the vector field in Eq.~\eqref{Asol} is roughly exponential ($A_+\sim e^{\pi \xi}$), variation of $\xi$ of order $2.5\times 10^{-3}$ have an impact below the percent level in the vector field. 

\section{Jordan and Einstein Frames}\label{AB}
Here we briefly discuss the transformation from Jordan to Einstein frames used in section~\ref{jtoe} applied to a single pseudoscalar field non-minimally coupled with gravity. Further details of the relation between Jordan and Einstein frames can be found for instance in Ref.~\cite{Kaiser:2010ps}. We start with the action for the system in Eq.~\eqref{lsgc}. By doing the conformal transformation $\bar{g}_{\mu\nu} = \Omega(\phi)  g_{\mu\nu}$, the action  (\ref{lsgc}) is transformed into 
\ba\label{lsgcef} 
{\cal L} &=& \sqrt{-\bar{g}} \left[  \frac{{M_P^{2}}}{2}  \left(  \bar{R} +\frac{3 \Box \Omega}{\Omega^2} -\frac{3 }{2 \Omega^2} \bar{g}^{\mu\nu} \bar{\nabla}_{\mu} \Omega \bar{\nabla}_{\nu} \Omega \right)  -  \frac{1}{2\Omega} (\bar{\nabla }\phi)^2- \bar{V}(\phi) \right. \nonumber\\ 
 &&\qquad\qquad- \left. \frac{\mathcal{N}}{4 \Omega^2} \left( F^{\mu \nu}F_{\mu \nu} + \frac{\alpha}{f }\phi  F^{\mu \nu}\tilde{F}_{\mu \nu}  \right)  \right],
\ea
where $\bar{V} (\phi) \equiv \frac{V(\phi)}{\Omega^2}$ and we have used the fact that the spacetime coordinates are not affected by the conformal transformation: $\partial_{\mu}  = \bar{\partial}_{\mu}$. Moreover, as the derivatives act on pseudoscalar fields, we can write $\bar{\nabla}_{\mu} \Omega = \bar{\partial}_{\mu} \Omega = \nabla_{\mu} \Omega= \partial_{\mu} \Omega$, where the covariant derivative $\bar{\nabla}_{\mu}$ is compatible with the metric $\bar{g}_{\mu\nu}$. The vector part is not altered since the field strength $F_{\mu \nu}$ is not affected by the conformal transformation. Now, recalling that $\Box \Omega = \frac{1}{\sqrt{-g}} \partial_{\mu} (\sqrt{-g} g^{\mu \nu} \partial_{\nu} \Omega)$ and that $\sqrt{-\bar{g} } = \Omega^2 \sqrt{-g}$, we realize that
\be
\sqrt{-\bar{g}}  \frac{{M_P^{2}}}{2} \frac{3 \Box \Omega}{\Omega^2}
\ee
is a boundary term. Additionally, the derivatives of the function $\Omega$ read
\be
\bar{g}^{\mu \nu} \bar{\nabla}_{\mu} \Omega \bar{\nabla}_{\nu} \Omega = \left(\frac{\partial \Omega}{\partial \phi} \right)^2  \bar{g}^{\mu \nu} \bar{\nabla}_{\mu} \phi \bar{\nabla}_{\nu} \phi\,,
\ee
and then, the kinetic term can be rewritten as:
\be
 \frac{3 M_P^{2} }{ 4 \Omega^2}  (\bar{\nabla} \Omega)^2  +\frac{1}{2\Omega} (\bar{\nabla }\phi)^2  = \frac{1}{2}K(\phi) \bar{g}^{\mu \nu} \bar{\nabla}_{\mu} \phi \bar{\nabla}_{\nu} \phi
\ee
where
\be\label{Kdef}
K(\phi)  =   \frac{1}{\Omega} + \frac{3 M_P^2}{2\Omega^2} \left( \frac{\partial \Omega}{\partial \phi} \right)^2.
\ee
So, the Lagrangian (\ref{lsgcef})  in the Einstein frame reduces to
\be\label{eomefapp}
{\cal L} = \sqrt{-\bar{g}} \left[  \frac{{M_P^{2}}}{2}    \bar{R}  -  \frac{1}{2}K(\phi) \bar{g}^{\mu \nu} \bar{\nabla}_{\mu} \phi \bar{\nabla}_{\nu} \phi -  \bar{V}(\phi)  - \frac{\mathcal{N}}{4 } \bar{F}^{\mu \nu}\bar{F}_{\mu \nu} - \frac{\mathcal{N}\alpha }{4f }\phi    \bar{F}^{\mu \nu}\tilde{\bar{F}}_{\mu \nu}  \right]. 
\ee
In the case of a single field, we can always define a canonical field $\bar{\phi}$ through the transformation $d \bar{\phi} /d \phi  = K ^{1/2}$. The action for the system becomes
\be\label{eomefcanapp}
{\cal L} = \sqrt{-\bar{g}} \left[  \frac{{M_P^{2}}}{2}    \bar{R}  -  \frac{1}{2}\bar{g}^{\mu \nu} \bar{\nabla}_{\mu} \bar{\phi} \bar{\nabla}_{\nu} \bar{\phi} -  \bar{V}(\bar{\phi})  - \frac{\mathcal{N}}{4 } \bar{F}^{\mu \nu}\bar{F}_{\mu \nu} -  \frac{\mathcal{N}\alpha}{4f }  \phi(\bar{\phi})    \bar{F}^{\mu \nu}\tilde{\bar{F}}_{\mu \nu}  \right],
\ee
which is the action for a canonical pseudoscalar field minimally coupled with gravity.

\bibliographystyle{JHEP} 
\bibliography{bibli} 

\providecommand{\href}[2]{#2}\begingroup\raggedright\begin{thebibliography}{10}

\bibitem{Anber:2009ua}
M.~M. Anber and L.~Sorbo, \emph{{Naturally inflating on steep potentials
  through electromagnetic dissipation}},
  \href{https://doi.org/10.1103/PhysRevD.81.043534}{\emph{Phys. Rev.}
  {\bfseries D81} (2010) 043534}
  [\href{https://arxiv.org/abs/0908.4089}{{\ttfamily 0908.4089}}].

\bibitem{Anber:2012du}
M.~M. Anber and L.~Sorbo, \emph{{Non-Gaussianities and chiral gravitational
  waves in natural steep inflation}},
  \href{https://doi.org/10.1103/PhysRevD.85.123537}{\emph{Phys. Rev.}
  {\bfseries D85} (2012) 123537}
  [\href{https://arxiv.org/abs/1203.5849}{{\ttfamily 1203.5849}}].

\bibitem{Barnaby:2010vf}
N.~Barnaby and M.~Peloso, \emph{{Large Nongaussianity in Axion Inflation}},
  \href{https://doi.org/10.1103/PhysRevLett.106.181301}{\emph{Phys. Rev. Lett.}
  {\bfseries 106} (2011) 181301}
  [\href{https://arxiv.org/abs/1011.1500}{{\ttfamily 1011.1500}}].

\bibitem{Sorbo:2011rz}
L.~Sorbo, \emph{{Parity violation in the Cosmic Microwave Background from a
  pseudoscalar inflaton}},
  \href{https://doi.org/10.1088/1475-7516/2011/06/003}{\emph{JCAP} {\bfseries
  1106} (2011) 003} [\href{https://arxiv.org/abs/1101.1525}{{\ttfamily
  1101.1525}}].

\bibitem{Barnaby:2011vw}
N.~Barnaby, R.~Namba and M.~Peloso, \emph{{Phenomenology of a Pseudo-Scalar
  Inflaton: Naturally Large Nongaussianity}},
  \href{https://doi.org/10.1088/1475-7516/2011/04/009}{\emph{JCAP} {\bfseries
  1104} (2011) 009} [\href{https://arxiv.org/abs/1102.4333}{{\ttfamily
  1102.4333}}].

\bibitem{Jimenez:2017cdr}
D.~Jim\'enez, K.~Kamada, K.~Schmitz and X.-J. Xu, \emph{{Baryon asymmetry and
  gravitational waves from pseudoscalar inflation}},
  \href{https://doi.org/10.1088/1475-7516/2017/12/011}{\emph{JCAP} {\bfseries
  1712} (2017) 011} [\href{https://arxiv.org/abs/1707.07943}{{\ttfamily
  1707.07943}}].

\bibitem{Shiraishi:2013kxa}
M.~Shiraishi, A.~Ricciardone and S.~Saga, \emph{{Parity violation in the CMB
  bispectrum by a rolling pseudoscalar}},
  \href{https://doi.org/10.1088/1475-7516/2013/11/051}{\emph{JCAP} {\bfseries
  1311} (2013) 051} [\href{https://arxiv.org/abs/1308.6769}{{\ttfamily
  1308.6769}}].

\bibitem{Bartolo:2014hwa}
N.~Bartolo, S.~Matarrese, M.~Peloso and M.~Shiraishi, \emph{{Parity-violating
  and anisotropic correlations in pseudoscalar inflation}},
  \href{https://doi.org/10.1088/1475-7516/2015/01/027}{\emph{JCAP} {\bfseries
  1501} (2015) 027} [\href{https://arxiv.org/abs/1411.2521}{{\ttfamily
  1411.2521}}].

\bibitem{Durrer:2010mq}
R.~Durrer, L.~Hollenstein and R.~K. Jain, \emph{{Can slow roll inflation induce
  relevant helical magnetic fields?}},
  \href{https://doi.org/10.1088/1475-7516/2011/03/037}{\emph{JCAP} {\bfseries
  1103} (2011) 037} [\href{https://arxiv.org/abs/1005.5322}{{\ttfamily
  1005.5322}}].

\bibitem{Caprini:2014mja}
C.~Caprini and L.~Sorbo, \emph{{Adding helicity to inflationary
  magnetogenesis}},
  \href{https://doi.org/10.1088/1475-7516/2014/10/056}{\emph{JCAP} {\bfseries
  1410} (2014) 056} [\href{https://arxiv.org/abs/1407.2809}{{\ttfamily
  1407.2809}}].

\bibitem{Garcia-Bellido:2016dkw}
J.~Garc\'ia-Bellido, M.~Peloso and C.~Unal, \emph{{Gravitational waves at
  interferometer scales and primordial black holes in axion inflation}},
  \href{https://doi.org/10.1088/1475-7516/2016/12/031}{\emph{JCAP} {\bfseries
  1612} (2016) 031} [\href{https://arxiv.org/abs/1610.03763}{{\ttfamily
  1610.03763}}].

\bibitem{Ferreira:2014zia}
R.~Z. Ferreira and M.~S. Sloth, \emph{{Universal Constraints on Axions from
  Inflation}}, \href{https://doi.org/10.1007/JHEP12(2014)139}{\emph{JHEP}
  {\bfseries 12} (2014) 139} [\href{https://arxiv.org/abs/1409.5799}{{\ttfamily
  1409.5799}}].

\bibitem{Ferreira:2015omg}
R.~Z. Ferreira, J.~Ganc, J.~Nore\~na and M.~S. Sloth, \emph{{On the validity of
  the perturbative description of axions during inflation}},
  \href{https://doi.org/10.1088/1475-7516/2016/04/039
  10.1088/1475-7516/2016/10/E01, 10.1088/1475-7516/2016/10/E01,
  10.1088/1475-7516/2016/04/039}{\emph{JCAP} {\bfseries 1604} (2016) 039}
  [\href{https://arxiv.org/abs/1512.06116}{{\ttfamily 1512.06116}}].

\bibitem{Peloso:2016gqs}
M.~Peloso, L.~Sorbo and C.~Unal, \emph{{Rolling axions during inflation:
  perturbativity and signatures}},
  \href{https://doi.org/10.1088/1475-7516/2016/09/001}{\emph{JCAP} {\bfseries
  1609} (2016) 001} [\href{https://arxiv.org/abs/1606.00459}{{\ttfamily
  1606.00459}}].

\bibitem{Anber:2006xt}
M.~M. Anber and L.~Sorbo, \emph{{N-flationary magnetic fields}},
  \href{https://doi.org/10.1088/1475-7516/2006/10/018}{\emph{JCAP} {\bfseries
  0610} (2006) 018} [\href{https://arxiv.org/abs/astro-ph/0606534}{{\ttfamily
  astro-ph/0606534}}].

\bibitem{Barnaby:2011qe}
N.~Barnaby, E.~Pajer and M.~Peloso, \emph{{Gauge Field Production in Axion
  Inflation: Consequences for Monodromy, non-Gaussianity in the CMB, and
  Gravitational Waves at Interferometers}},
  \href{https://doi.org/10.1103/PhysRevD.85.023525}{\emph{Phys. Rev.}
  {\bfseries D85} (2012) 023525}
  [\href{https://arxiv.org/abs/1110.3327}{{\ttfamily 1110.3327}}].

\bibitem{Barnaby:2012tk}
N.~Barnaby, R.~Namba and M.~Peloso, \emph{{Observable non-gaussianity from
  gauge field production in slow roll inflation, and a challenging connection
  with magnetogenesis}},
  \href{https://doi.org/10.1103/PhysRevD.85.123523}{\emph{Phys. Rev.}
  {\bfseries D85} (2012) 123523}
  [\href{https://arxiv.org/abs/1202.1469}{{\ttfamily 1202.1469}}].

\bibitem{Dimopoulos:2012av}
K.~Dimopoulos and M.~Karciauskas, \emph{{Parity Violating Statistical
  Anisotropy}}, \href{https://doi.org/10.1007/JHEP06(2012)040}{\emph{JHEP}
  {\bfseries 06} (2012) 040} [\href{https://arxiv.org/abs/1203.0230}{{\ttfamily
  1203.0230}}].

\bibitem{Meerburg:2012id}
P.~D. Meerburg and E.~Pajer, \emph{{Observational Constraints on Gauge Field
  Production in Axion Inflation}},
  \href{https://doi.org/10.1088/1475-7516/2013/02/017}{\emph{JCAP} {\bfseries
  1302} (2013) 017} [\href{https://arxiv.org/abs/1203.6076}{{\ttfamily
  1203.6076}}].

\bibitem{Barnaby:2012xt}
N.~Barnaby, J.~Moxon, R.~Namba, M.~Peloso, G.~Shiu and P.~Zhou, \emph{{Gravity
  waves and non-Gaussian features from particle production in a sector
  gravitationally coupled to the inflaton}},
  \href{https://doi.org/10.1103/PhysRevD.86.103508}{\emph{Phys. Rev.}
  {\bfseries D86} (2012) 103508}
  [\href{https://arxiv.org/abs/1206.6117}{{\ttfamily 1206.6117}}].

\bibitem{Linde:2012bt}
A.~Linde, S.~Mooij and E.~Pajer, \emph{{Gauge field production in supergravity
  inflation: Local non-Gaussianity and primordial black holes}},
  \href{https://doi.org/10.1103/PhysRevD.87.103506}{\emph{Phys. Rev.}
  {\bfseries D87} (2013) 103506}
  [\href{https://arxiv.org/abs/1212.1693}{{\ttfamily 1212.1693}}].

\bibitem{Cook:2013xea}
J.~L. Cook and L.~Sorbo, \emph{{An inflationary model with small scalar and
  large tensor nongaussianities}},
  \href{https://doi.org/10.1088/1475-7516/2013/11/047}{\emph{JCAP} {\bfseries
  1311} (2013) 047} [\href{https://arxiv.org/abs/1307.7077}{{\ttfamily
  1307.7077}}].

\bibitem{Fleury:2014qfa}
P.~Fleury, J.~P. Beltr\'an~Almeida, C.~Pitrou and J.-P. Uzan, \emph{{On the
  stability and causality of scalar-vector theories}},
  \href{https://doi.org/10.1088/1475-7516/2014/11/043}{\emph{JCAP} {\bfseries
  1411} (2014) 043} [\href{https://arxiv.org/abs/1406.6254}{{\ttfamily
  1406.6254}}].

\bibitem{Bartolo:2015dga}
N.~Bartolo, S.~Matarrese, M.~Peloso and M.~Shiraishi, \emph{{Parity-violating
  CMB correlators with non-decaying statistical anisotropy}},
  \href{https://doi.org/10.1088/1475-7516/2015/07/039}{\emph{JCAP} {\bfseries
  1507} (2015) 039} [\href{https://arxiv.org/abs/1505.02193}{{\ttfamily
  1505.02193}}].

\bibitem{Namba:2015gja}
R.~Namba, M.~Peloso, M.~Shiraishi, L.~Sorbo and C.~Unal, \emph{{Scale-dependent
  gravitational waves from a rolling axion}},
  \href{https://doi.org/10.1088/1475-7516/2016/01/041}{\emph{JCAP} {\bfseries
  1601} (2016) 041} [\href{https://arxiv.org/abs/1509.07521}{{\ttfamily
  1509.07521}}].

\bibitem{Domcke:2016bkh}
V.~Domcke, M.~Pieroni and P.~Bin\'etruy, \emph{{Primordial gravitational waves
  for universality classes of pseudoscalar inflation}},
  \href{https://doi.org/10.1088/1475-7516/2016/06/031}{\emph{JCAP} {\bfseries
  1606} (2016) 031} [\href{https://arxiv.org/abs/1603.01287}{{\ttfamily
  1603.01287}}].

\bibitem{Shiraishi:2016yun}
M.~Shiraishi, C.~Hikage, R.~Namba, T.~Namikawa and M.~Hazumi, \emph{{Testing
  statistics of the CMB B-mode polarization toward unambiguously establishing
  quantum fluctuation of the vacuum}},
  \href{https://doi.org/10.1103/PhysRevD.94.043506}{\emph{Phys. Rev.}
  {\bfseries D94} (2016) 043506}
  [\href{https://arxiv.org/abs/1606.06082}{{\ttfamily 1606.06082}}].

\bibitem{Almeida:2017lrq}
J.~P. Beltr\'an~Almeida, J.~Motoa-Manzano and C.~A. Valenzuela-Toledo,
  \emph{{de Sitter symmetries and inflationary correlators in parity violating
  scalar-vector models}},
  \href{https://doi.org/10.1088/1475-7516/2017/11/015}{\emph{JCAP} {\bfseries
  1711} (2017) 015} [\href{https://arxiv.org/abs/1706.05099}{{\ttfamily
  1706.05099}}].

\bibitem{Caprini:2017vnn}
C.~Caprini, M.~C. Guzzetti and L.~Sorbo, \emph{{Inflationary magnetogenesis
  with added helicity: constraints from non-Gaussianities}},
  \href{https://doi.org/10.1088/1361-6382/aac143}{\emph{Class. Quant. Grav.}
  {\bfseries 35} (2018) 124003}
  [\href{https://arxiv.org/abs/1707.09750}{{\ttfamily 1707.09750}}].

\bibitem{Adshead:2018oaa}
P.~Adshead, L.~Pearce, M.~Peloso, M.~A. Roberts and L.~Sorbo,
  \emph{{Phenomenology of fermion production during axion inflation}},
  \href{https://arxiv.org/abs/1803.04501}{{\ttfamily 1803.04501}}.

\bibitem{Futamase:1987ua}
T.~Futamase and K.-i. Maeda, \emph{{Chaotic Inflationary Scenario in Models
  Having Nonminimal Coupling With Curvature}},
  \href{https://doi.org/10.1103/PhysRevD.39.399}{\emph{Phys. Rev.} {\bfseries
  D39} (1989) 399}.

\bibitem{Fakir:1990eg}
R.~Fakir and W.~G. Unruh, \emph{{Improvement on cosmological chaotic inflation
  through nonminimal coupling}},
  \href{https://doi.org/10.1103/PhysRevD.41.1783}{\emph{Phys. Rev.} {\bfseries
  D41} (1990) 1783}.

\bibitem{Makino:1991sg}
N.~Makino and M.~Sasaki, \emph{{The Density perturbation in the chaotic
  inflation with nonminimal coupling}},
  \href{https://doi.org/10.1143/PTP.86.103}{\emph{Prog. Theor. Phys.}
  {\bfseries 86} (1991) 103}.

\bibitem{Muta:1991mw}
T.~Muta and S.~D. Odintsov, \emph{{Model dependence of the nonminimal scalar
  graviton effective coupling constant in curved space-time}},
  \href{https://doi.org/10.1142/S0217732391004206}{\emph{Mod. Phys. Lett.}
  {\bfseries A6} (1991) 3641}.

\bibitem{Mukaigawa:1997nh}
S.~Mukaigawa, T.~Muta and S.~D. Odintsov, \emph{{Finite grand unified theories
  and inflation}}, \href{https://doi.org/10.1142/S0217751X98001396}{\emph{Int.
  J. Mod. Phys.} {\bfseries A13} (1998) 2739}
  [\href{https://arxiv.org/abs/hep-ph/9709299}{{\ttfamily hep-ph/9709299}}].

\bibitem{Komatsu:1997hv}
E.~Komatsu and T.~Futamase, \emph{{Constraints on the chaotic inflationary
  scenario with a nonminimally coupled `inflaton' field from the cosmic
  microwave background radiation anisotropy}},
  \href{https://doi.org/10.1103/PhysRevD.58.023004,
  10.1103/PhysRevD.58.089902}{\emph{Phys. Rev.} {\bfseries D58} (1998) 023004}
  [\href{https://arxiv.org/abs/astro-ph/9711340}{{\ttfamily
  astro-ph/9711340}}].

\bibitem{Komatsu:1999mt}
E.~Komatsu and T.~Futamase, \emph{{Complete constraints on a nonminimally
  coupled chaotic inflationary scenario from the cosmic microwave background}},
  \href{https://doi.org/10.1103/PhysRevD.59.064029}{\emph{Phys. Rev.}
  {\bfseries D59} (1999) 064029}
  [\href{https://arxiv.org/abs/astro-ph/9901127}{{\ttfamily
  astro-ph/9901127}}].

\bibitem{Bezrukov:2007ep}
F.~L. Bezrukov and M.~Shaposhnikov, \emph{{The Standard Model Higgs boson as
  the inflaton}},
  \href{https://doi.org/10.1016/j.physletb.2007.11.072}{\emph{Phys. Lett.}
  {\bfseries B659} (2008) 703}
  [\href{https://arxiv.org/abs/0710.3755}{{\ttfamily 0710.3755}}].

\bibitem{GarciaBellido:2011de}
J.~Garc\'ia-Bellido, J.~Rubio, M.~Shaposhnikov and D.~Zenhausern,
  \emph{{Higgs-Dilaton Cosmology: From the Early to the Late Universe}},
  \href{https://doi.org/10.1103/PhysRevD.84.123504}{\emph{Phys. Rev.}
  {\bfseries D84} (2011) 123504}
  [\href{https://arxiv.org/abs/1107.2163}{{\ttfamily 1107.2163}}].

\bibitem{Kallosh:2013hoa}
R.~Kallosh and A.~Linde, \emph{{Universality Class in Conformal Inflation}},
  \href{https://doi.org/10.1088/1475-7516/2013/07/002}{\emph{JCAP} {\bfseries
  1307} (2013) 002} [\href{https://arxiv.org/abs/1306.5220}{{\ttfamily
  1306.5220}}].

\bibitem{Kallosh:2013maa}
R.~Kallosh and A.~Linde, \emph{{Non-minimal Inflationary Attractors}},
  \href{https://doi.org/10.1088/1475-7516/2013/10/033}{\emph{JCAP} {\bfseries
  1310} (2013) 033} [\href{https://arxiv.org/abs/1307.7938}{{\ttfamily
  1307.7938}}].

\bibitem{Kallosh:2013tua}
R.~Kallosh, A.~Linde and D.~Roest, \emph{{Universal Attractor for Inflation at
  Strong Coupling}},
  \href{https://doi.org/10.1103/PhysRevLett.112.011303}{\emph{Phys. Rev. Lett.}
  {\bfseries 112} (2014) 011303}
  [\href{https://arxiv.org/abs/1310.3950}{{\ttfamily 1310.3950}}].

\bibitem{Kallosh:2013yoa}
R.~Kallosh, A.~Linde and D.~Roest, \emph{{Superconformal Inflationary
  $\alpha$-Attractors}},
  \href{https://doi.org/10.1007/JHEP11(2013)198}{\emph{JHEP} {\bfseries 11}
  (2013) 198} [\href{https://arxiv.org/abs/1311.0472}{{\ttfamily 1311.0472}}].

\bibitem{Racioppi:2018zoy}
A.~Racioppi, \emph{{A new universal attractor: linear inflation}},
  \href{https://arxiv.org/abs/1801.08810}{{\ttfamily 1801.08810}}.

\bibitem{Kaiser:2010yu}
D.~I. Kaiser and A.~T. Todhunter, \emph{{Primordial Perturbations from
  Multifield Inflation with Nonminimal Couplings}},
  \href{https://doi.org/10.1103/PhysRevD.81.124037}{\emph{Phys. Rev.}
  {\bfseries D81} (2010) 124037}
  [\href{https://arxiv.org/abs/1004.3805}{{\ttfamily 1004.3805}}].

\bibitem{Greenwood:2012aj}
R.~N. Greenwood, D.~I. Kaiser and E.~I. Sfakianakis, \emph{{Multifield Dynamics
  of Higgs Inflation}},
  \href{https://doi.org/10.1103/PhysRevD.87.064021}{\emph{Phys. Rev.}
  {\bfseries D87} (2013) 064021}
  [\href{https://arxiv.org/abs/1210.8190}{{\ttfamily 1210.8190}}].

\bibitem{Kaiser:2013sna}
D.~I. Kaiser and E.~I. Sfakianakis, \emph{{Multifield Inflation after Planck:
  The Case for Nonminimal Couplings}},
  \href{https://doi.org/10.1103/PhysRevLett.112.011302}{\emph{Phys. Rev. Lett.}
  {\bfseries 112} (2014) 011302}
  [\href{https://arxiv.org/abs/1304.0363}{{\ttfamily 1304.0363}}].

\bibitem{Tenkanen:2017jih}
T.~Tenkanen, \emph{{Resurrecting Quadratic Inflation with a non-minimal
  coupling to gravity}},
  \href{https://doi.org/10.1088/1475-7516/2017/12/001}{\emph{JCAP} {\bfseries
  1712} (2017) 001} [\href{https://arxiv.org/abs/1710.02758}{{\ttfamily
  1710.02758}}].

\bibitem{Markkanen:2017tun}
T.~Markkanen, T.~Tenkanen, V.~Vaskonen and H.~Veermae, \emph{{Quantum
  corrections to quartic inflation with a non-minimal coupling: metric vs.
  Palatini}}, \href{https://doi.org/10.1088/1475-7516/2018/03/029}{\emph{JCAP}
  {\bfseries 1803} (2018) 029}
  [\href{https://arxiv.org/abs/1712.04874}{{\ttfamily 1712.04874}}].

\bibitem{Jarv:2017azx}
L.~Jarv, A.~Racioppi and T.~Tenkanen, \emph{{Palatini side of inflationary
  attractors}}, \href{https://doi.org/10.1103/PhysRevD.97.083513}{\emph{Phys.
  Rev.} {\bfseries D97} (2018) 083513}
  [\href{https://arxiv.org/abs/1712.08471}{{\ttfamily 1712.08471}}].

\bibitem{Domcke:2017fix}
V.~Domcke, F.~Muia, M.~Pieroni and L.~T. Witkowski, \emph{{PBH dark matter from
  axion inflation}},
  \href{https://doi.org/10.1088/1475-7516/2017/07/048}{\emph{JCAP} {\bfseries
  1707} (2017) 048} [\href{https://arxiv.org/abs/1704.03464}{{\ttfamily
  1704.03464}}].

\bibitem{Wu:2016hul}
W.~L.~K. Wu et~al., \emph{{Initial Performance of BICEP3: A Degree Angular
  Scale 95 GHz Band Polarimeter}},
  \href{https://doi.org/10.1007/s10909-015-1403-x}{\emph{J. Low. Temp. Phys.}
  {\bfseries 184} (2016) 765}
  [\href{https://arxiv.org/abs/1601.00125}{{\ttfamily 1601.00125}}].

\bibitem{Matsumura:2013aja}
T.~Matsumura et~al., \emph{{Mission design of LiteBIRD}},
  \href{https://arxiv.org/abs/1311.2847}{{\ttfamily 1311.2847}}.

\bibitem{Ade:2018sbj}
{\scshape Simons Observatory} collaboration, J.~Aguirre et~al., \emph{{The
  Simons Observatory: Science goals and forecasts}},
  \href{https://arxiv.org/abs/1808.07445}{{\ttfamily 1808.07445}}.

\bibitem{Ade:2015lrj}
{\scshape Planck} collaboration, P.~A.~R. Ade et~al., \emph{{Planck 2015
  results. XX. Constraints on inflation}},
  \href{https://doi.org/10.1051/0004-6361/201525898}{\emph{Astron. Astrophys.}
  {\bfseries 594} (2016) A20}
  [\href{https://arxiv.org/abs/1502.02114}{{\ttfamily 1502.02114}}].

\bibitem{Freese:1990rb}
K.~Freese, J.~A. Frieman and A.~V. Olinto, \emph{{Natural inflation with pseudo
  - Nambu-Goldstone bosons}},
  \href{https://doi.org/10.1103/PhysRevLett.65.3233}{\emph{Phys. Rev. Lett.}
  {\bfseries 65} (1990) 3233}.

\bibitem{Adams:1992bn}
F.~C. Adams, J.~R. Bond, K.~Freese, J.~A. Frieman and A.~V. Olinto,
  \emph{{Natural inflation: Particle physics models, power law spectra for
  large scale structure, and constraints from COBE}},
  \href{https://doi.org/10.1103/PhysRevD.47.426}{\emph{Phys. Rev.} {\bfseries
  D47} (1993) 426} [\href{https://arxiv.org/abs/hep-ph/9207245}{{\ttfamily
  hep-ph/9207245}}].

\bibitem{Ade:2015ava}
{\scshape Planck} collaboration, P.~A.~R. Ade et~al., \emph{{Planck 2015
  results. XVII. Constraints on primordial non-Gaussianity}},
  \href{https://doi.org/10.1051/0004-6361/201525836}{\emph{Astron. Astrophys.}
  {\bfseries 594} (2016) A17}
  [\href{https://arxiv.org/abs/1502.01592}{{\ttfamily 1502.01592}}].

\bibitem{Kaiser:2010ps}
D.~I. Kaiser, \emph{{Conformal Transformations with Multiple Scalar Fields}},
  \href{https://doi.org/10.1103/PhysRevD.81.084044}{\emph{Phys. Rev.}
  {\bfseries D81} (2010) 084044}
  [\href{https://arxiv.org/abs/1003.1159}{{\ttfamily 1003.1159}}].

\bibitem{Kamenshchik:2014waa}
A.~{\relax Yu}. Kamenshchik and C.~F. Steinwachs, \emph{{Question of quantum
  equivalence between Jordan frame and Einstein frame}},
  \href{https://doi.org/10.1103/PhysRevD.91.084033}{\emph{Phys. Rev.}
  {\bfseries D91} (2015) 084033}
  [\href{https://arxiv.org/abs/1408.5769}{{\ttfamily 1408.5769}}].

\bibitem{Germani:2010hd}
C.~Germani and A.~Kehagias, \emph{{UV-Protected Inflation}},
  \href{https://doi.org/10.1103/PhysRevLett.106.161302}{\emph{Phys. Rev. Lett.}
  {\bfseries 106} (2011) 161302}
  [\href{https://arxiv.org/abs/1012.0853}{{\ttfamily 1012.0853}}].

\bibitem{Folkerts:2013tua}
S.~Folkerts, C.~Germani and J.~Redondo, \emph{{Axion Dark Matter and Planck
  favor non-minimal couplings to gravity}},
  \href{https://doi.org/10.1016/j.physletb.2013.12.026}{\emph{Phys. Lett.}
  {\bfseries B728} (2014) 532}
  [\href{https://arxiv.org/abs/1304.7270}{{\ttfamily 1304.7270}}].

\bibitem{Chiba:2008ia}
T.~Chiba and M.~Yamaguchi, \emph{{Extended Slow-Roll Conditions and Rapid-Roll
  Conditions}},
  \href{https://doi.org/10.1088/1475-7516/2008/10/021}{\emph{JCAP} {\bfseries
  0810} (2008) 021} [\href{https://arxiv.org/abs/0807.4965}{{\ttfamily
  0807.4965}}].

\bibitem{Nozari:2015ada}
K.~Nozari, H.~Kahvaee and N.~Rashidi, \emph{{Consistency relation for natural
  inflation with Planck 2015 data}},
  \href{https://doi.org/10.1007/s10509-015-2514-1}{\emph{Astrophys. Space Sci.}
  {\bfseries 359} (2015) 63}.

\end{thebibliography}\endgroup

\end{document}